\documentclass[%
reprint,
 amsmath,amssymb,
 aps,
floatfix,
]{revtex4-2}

\usepackage{hyperref}
\usepackage{cleveref}
\usepackage{graphicx}
\usepackage{enumerate} 
\usepackage[tight,TABTOPCAP]{subfigure}
\usepackage{amssymb}
\usepackage{setspace}
\usepackage{xcolor}
\newcommand{\change}[1]{#1}
\Crefname{equation}{Eq.~}{Eqs.~}
\Crefname{figure}{Fig.~}{Figs.~}
\Crefname{section}{Sec.~}{Secs.~}
\usepackage{tikz}
\usetikzlibrary{shapes.geometric, arrows}

\tikzstyle{startstop} = [rectangle, rounded corners, minimum width=3cm, minimum height=1cm,text centered, draw=black, fill=red!30]

\tikzstyle{io} = [rectangle, rounded corners, minimum width=3cm, minimum height=1cm,text centered, draw=black, fill=blue!30]

\tikzstyle{method} = [rectangle, rounded corners, minimum width=3cm, minimum height=1cm,text centered, draw=black, fill=green!30]

\tikzstyle{arrow} = [thick,->,>=stealth]

\usepackage{soul}

\begin{document}

\title
  {Data-driven identification and analysis of the glass transition in polymer melts}

\author{Atreyee Banerjee}

\author{Hsiao-Ping Hsu}

\author{Kurt Kremer}

\author{Oleksandra Kukharenko}
\email{kukharenko@mpip-mainz.mpg.de}
\affiliation
{Theory Department, Max Planck Institute for Polymer Research, \\Ackermannweg 10, 55128 Mainz, Germany}

\title[Data-driven identification of glass transition T]
  {Data-driven identification and analysis of the glass transition in polymer melts}






\begin{abstract}

Understanding the nature of glass transition, as well as precise estimation of the glass transition temperature for polymeric materials, remain open questions in both experimental and theoretical polymer sciences. We propose a data-driven approach, which utilizes the high-resolution details accessible through the molecular dynamics simulation and considers the structural information of individual chains. It clearly identifies the glass transition temperature of polymer melts of weakly semiflexible chains. By combining principal component analysis and clustering, we identify the glass transition temperature in the asymptotic limit even from relatively short-time trajectories, which just reach into the Rouse-like monomer displacement regime. We demonstrate that fluctuations captured by the principal component analysis reflect the change in a chain's behaviour: from conformational rearrangement above to small rearrangements below the glass transition temperature. Our approach is straightforward to apply, and should be applicable to other polymeric glass-forming liquids.

\end{abstract}


\maketitle

 Polymer materials in applications are often in the glassy state. Upon cooling of a rubbery liquid polymer, dynamic properties such as viscosity or relaxation time increase drastically near the glass transition temperature ($T_g$) in a super-Arrhenius fashion~\cite{angell1988perspective,angell1995formation,binder2003glass,berthier2011theoretical} without any remarkable change in structural properties~\cite{binder2003glass}. Despite enormous  experimental and theoretical efforts~\cite{kirkpatrick1989scaling,gotze1992relaxation,cugliandolo1993analytical,wolynes1997entropy,sastry1998signatures,dudowicz2005glass}, the nature of glass transition  as well as the question of a precisely defined $T_g$ still remain unclear~\cite{berthier2011theoretical,biroli2013perspective,lin2021glass,jin2022,godey2019local}. 
In computer simulations, $T_g$ is often calculated from characteristic macroscopic properties, e.g. changes in the specific volume, density or in energy~\cite{debenedetti2001supercooled,schnell2011simulated,godey2019local}. The increase in viscosity, equivalently of the terminal relaxation times, is commonly fitted to a Vogel-Fulcher-Tamann behavior which predicts a divergence at $T_{\rm{VFT}}$~\cite{soldera2006glass}, typically about $50^o$ below the calorimetric $T_g$~\cite{zhao2012temperature}.
However, the precise value of the observed $T_g$ depends on the cooling rate and fitting procedures, which can lead  to some ambiguities in comparison with experimental values~\cite{yasoshima2017diffusion,godey2018extent}
unlike a sharp and distinct change in physical properties.  
Thus reliable predictions of $T_g$ are indeed challenging ~\cite{lin2021glass,jin2022,chu2020understanding,deng2021role}.

Attempts to link $T_g$ with the molecular structure of polymeric materials draw more attention. Recent studies predict $T_g$ by quantifying the changes in specific dihedral angles and transitions between states defined by those angles \cite{godey2019local, jin2022} or by using averaged intra-chain properties \cite{baker2022cooperative}. 
A possibility to specify the structural properties of the glassy systems which can reflect changes in $T_g$ is attractive, but it remains challenging and system specific. 
Machine learning (ML) methods hold a great promise to automatize the determination of structural descriptors from molecular simulation data.
Recently, the application of ML to non-polymeric supercooled model liquids allowed to understand the connection between characteristic local structures and the slowing down of dynamical properties~\cite{Schoenholz2016,Kohli2020, Boattini2020, Boattini2021, Clegg2021, Coslovich2022}.
For polymer chains in a melt, the intra-chain properties associated with the chain connectivity and flexibility also play an important role in determining $T_g$. However, application of ML methods to determine structural changes during the glass transition in polymer chains is limited~\cite{Takano2017,Washizu2021,jin2022}. 

In this letter, we use unsupervised data-driven methods to identify the glass transition of polymer melts of weakly entangled polymer chains only by employing information about conformational fluctuations  at different  temperatures. 
We first analyse the combined data from different temperatures using principal component analysis (PCA)~\cite{abdi2010principal}, followed by clustering and determine a clear signature of glass transition. 
Considering the simulation data within a finite observation time window up into the Rouse-like regime, our approach allows a very solid extrapolation to infinite times to predict $T_g$. 
We then also employ the data-driven methods on individual temperature data separately. 
The non-monotonic variation of the magnitudes of leading eigenvalues and the participation ratio derived from PCA captures the signature of the glass transition. 
It also reflects a change in the nature of the fluctuations in the system.
We apply these approaches to the simulation data of a coarse-grained polymer model~\cite{Hsu2019} and compare estimates of $T_g$ obtained from classical fitting of macroscopic properties with the new method. The proposed method has the following advantages: (a) our approach is based on high-resolution microscopic details instead of average macroscopic properties, (b) it does not rely on the fitting protocols, (c) our analysis focuses on the information about structural fluctuations at the level of individual chains to predict $T_g$ from very moderate simulation trajectories.

\begin{figure*}
	\centering
        \centering
		\includegraphics[width=0.85\linewidth]{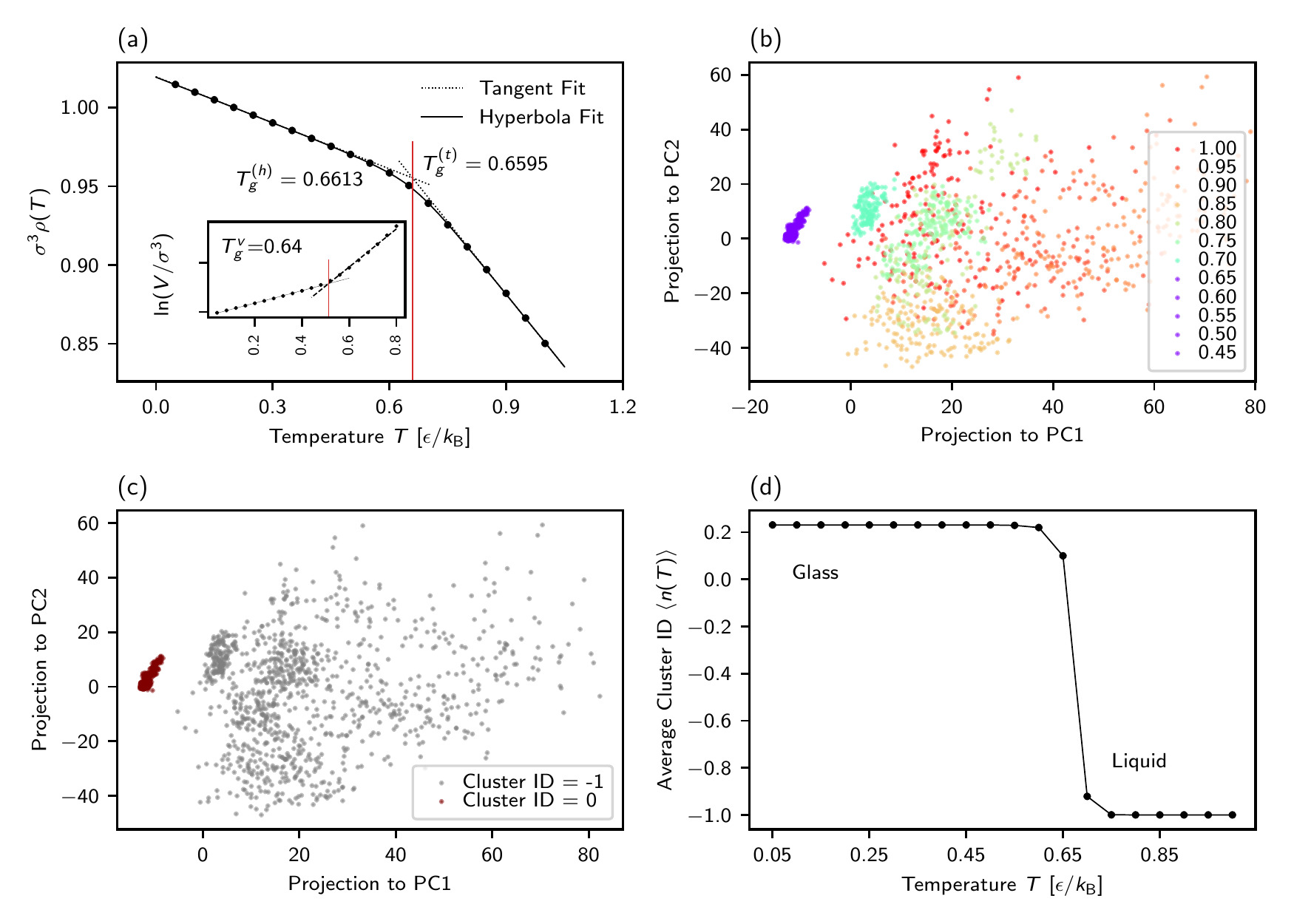}
		\caption{(a) Conventional methods of estimating the glass transition temperature $T_g$: Density $\rho(T)$ and logarithm of volume, $\ln(V/\sigma^3)$ (in the inset), plotted versus $T$. Estimates of $T_g$ via the two tangent (t) fits (dotted lines) at high and low $T$, hyperbola (h) fit (curve), and two linear fits (dashed lines in the inset) are indicated by vertical lines. (Data are taken from Ref.~\citenum{Hsu2019}).
		(b-d) Data-driven determination of $T_g$.
		(b) Projections of concatenated data from all $T$ for a single chain over multiple time frames in the two first leading PCs. 
		Each point in the plot corresponds to one chain's conformation at a given temperature at each time. Projections for $T>T_g$
		are colored varying from red to green while they are in purple for $T<T_g$ (data shown for $T \ge 0.45$ for clarity). 
		Note that the axis values in the PCA embedding don't correspond to a directly measurable physical quantity, rather could be viewed as a weighted linear combination of scaled input distances. 
  (c) DBSCAN of the PCA projection. The same projection as \Cref{fig:glass-transition-both}b colored with DBSCAN cluster indices (ID) instead of temperature. DBSCAN assigns the high-temperature liquid state as noise (cluster ID $=-1$) 
  and the low-temperature glassy state as a cluster (cluster ID $=0$).
		(d) Average cluster ID over all  chains versus $T$.
The separation between the liquid and glass state becomes sharper if we use median instead of mean (see SI, Fig. S6b).}
		\label{fig:glass-transition-both}
\end{figure*}

In Ref.~\citenum{Hsu2019}, Hsu and Kremer developed a new variant of bead-spring model~\cite{Kremer1990,Kremer1992} for studying the glass transition of polymer melts
~\cite{Xu2020,xu2020Role}.
Molecular dynamics simulations of a bulk polymer melt containing $n_c=2000$ semiflexible polymer chains of chain length $n_m=50$ monomers with Kuhn length $\approx 2.66\sigma$~\cite{Hsu2016} were first performed in the NPT ensemble at $P\approx 0\epsilon/\sigma^3$ and constant $T$ following a standard stepwise cooling protocol (20~temperatures from 1.0 to 0.05~$\epsilon/k_B$)
choosing a fast fixed cooling rate of $\Gamma = 8.3 \times 10^{-7} \epsilon/(k_B\tau)$ (see SI, Sec.~S-I for details).
The Rouse time is $\tau_R=\tau_0n_m^2 \approx 7225\tau$, and the entanglement time is $\tau_e=\tau_0 N_e^2 \approx 2266\tau$ with the characteristic relaxation time $\tau_0 \approx 2.89\tau$ estimated at $T=1.0\epsilon/k_B$, and the entanglement length $N_e = 28$ monomers~\cite{Hsu2016}. Here $\tau_0$ is the upper limit of time that a monomer can move freely. After the step cooling, subsequent NVT runs up to $3\times 10^4\tau$ were performed at each $T$ to investigate the monomer mobility characterized by the mean square displacement $g_1(t)$ (for details see Supporting Information (SI), Sec. S-I, Fig. S1). In this letter, we mainly use simulation trajectories from NVT runs stored every $200\tau$ in the time window between $200\tau$ and $3\times 10^4\tau$ (gray area in Fig. S1, SI) resulting in 150 frames per temperature. 

The first estimate of glass transition temperature at $T_g \approx 0.64 \epsilon/k_B$ using conventional fitting procedure was determined from the volume change (\Cref{fig:glass-transition-both}a inset).~\cite{Hsu2019} We here adapt another standard approach to estimate $T_g$ by performing a hyperbola fit~\cite{Patrone2016} on the temperature-dependent density of polymer melt  for $0.1\le k_BT/\epsilon \le 1.0$, $\rho(T)=c-a(T-T_0)-\frac{b}{2}(T-T_0+\sqrt{(T-T_0)^2+4e^f}$, where $c$, $T_0$, $a$, $b$ and $f$ are fitting parameters. $T_g$ is either defined by $T_g=T_0$ or the intersection point of the two tangents drawn at the high and the low temperature. Both give an identical, more precise estimate of $T_g = 0.660(4)\epsilon/k_B$ as shown in \Cref{fig:glass-transition-both}a and are used as reference values for evaluating the data-driven approach presented below. Note that $T_g$ obtained from the simulation data depends on cooling rate. 
We propose here an alternative data-driven approach to gain insight into glass transition  with a minimum a priori knowledge about the system and user input.

The analysis workflow consists of two different, but related methods (a sketch is given in SI, Sec. S-II). Both identify the same $T_g$, but treat the data differently (using combined information from all $20$ temperatures or individual information from each temperature).
To identify changes in the studied systems, we first define possible descriptors: sets of all pairwise internal distances for a single chain. They are well suited to describe conformational fluctuations of individual polymer chains. 
Then we apply principal component analysis (PCA)~\cite{abdi2010principal} to the high-dimensional descriptor space. PCA has been successfully used to  characterise the phase transition in conserved Ising spin systems~\cite{Wang2016,Wang2017}.
The method relies on purely structural information without any a priory knowledge of dynamical correlations.
A $M \times L$ real matrix $\mathbf{X}_c$ with elements $x^{c}_{m,l}$, $1 \leq m \leq M$, $1 \leq l \leq L$ is used to represent data for a single chain $c$. Here $c = 1,.., n_c$ is a chain index, $L$ is the number of descriptors (e.g.~the intra-chain distances between any two monomers in a single chain of $n_m=50$ monomers: $L = n_m\times(n_m-1)/2$ = 1225) 
and $M$ is the number of observations (i.e. $M=150 \textrm{ (time frames)} \times 20 \textrm{ (temperatures)}=3000$ for Method I, and $M=150 \textrm{ (time frames)} \times 1 \textrm{ (temperature)}=150$ for Method II).
$\mathbf{X}_c$ is standardised column-wise i.e. each element $x^c_{m,l}$ is converted to $\frac{x^c_{m,l}-\mu^c_l}{\sigma^c_l}$ , where $\mu^c_l = \frac{1}{M}\sum^M_{m=1}x^c_{m,l}$ is the mean value for each column $l$, and $\sigma^c_l = \sqrt{\frac{1}{M} \sum^M_{m=1} \left(x^c_{m,l} - \mu^c_l \right)^2}$ is its corresponding 
standard deviation for chain $c$ such that the rescaled columns $\textbf{x}_{l}^c$ have a mean value of 0 and a variance of $1$.
PCA is done individually for each chain by first calculating the covariance matrix $\mathbf{C}_c= \mathbf{X}_c ^T \mathbf{X}_c$, where $s^c_{j,k} = s^{c}_{k,j} = \frac{1}{M}\sum^M_{m=1}x^c_{m,j}x^c_{m,k} \ge 0$, $1\leq j,k \leq L$ are elements of $\mathbf{C}_c$ and $x^c_{m,j}$, $x^c_{m,K}$ are the standardised descriptors.
Then the eigenvalues $\lambda_{c,i}$ and corresponding eigenvectors $\mathbf{v}_{c,i}$ of the matrix $C_c$, for $i=1,2,3,\ldots,\textrm{min}(L,M)$ are calculated and sorted in decreasing order of $\lambda_{c,i}$. 
The original data set $\mathbf{X}_c$ is converted to $\tilde{\mathbf{X}}_{c}=\mathbf{X}_c \bf{v}_{c,k}$ by projecting $\mathbf{X}_c$ to the new orthogonal basis formed by $P$-leading eigenvectors $\mathbf{v}_{c,k}$,  where elements of $\tilde{\mathbf{X}}_{c}$ are $\tilde{x}^c_{k,m}=\sum^L_{l=1}x^c_{m,l}v^c_{k,l}$, $k=1,...,P$, and $P \le \textrm{min}(L,M)$ is the reduced number of dimensions ($P=4$ in this work).

Due to correlated motions of neighbouring monomers the intra-chain distance space can be reduced by skipping some distances. We discuss this in more detail in SI, Sec.~S-IX. All results are similar in nature after reducing the input feature space and  the asymptotic estimate of glass transition temperature is reported considering every fifth monomers in a chain.

\textit{Method I}. We perform PCA on a randomly selected single chain
using the internal distances over time concatenated for all temperatures. In this way we construct the new basis formed by eigenvectors $\mathbf{v}_{c,i}$ containing information about fluctuations of internal distances at all temperatures. The internal distances of the chain at each simulation snapshot and temperature are projected independently on this new basis.
Thus, projections in the new PCA space can be viewed as linear combinations of input distances. Already in two-dimensional projection, one could clearly differentiate between two states (\Cref{fig:glass-transition-both}b), which occur roughly around the glass transition temperature $T_g\approx 0.65\epsilon/k_B$ (\Cref{fig:glass-transition-both}a). 
 The scatter of the PCA projection qualitatively changes at and below $0.65 \epsilon/k_B$ indicating the onset of a different state.

To quantify the separation between liquid and glassy state, we perform such a PCA for each chain separately, followed by clustering. Clustering groups the chains' conformations at each simulation snapshot and temperature based on similarities in their conformational fluctuations reflected as closeness in the PCA projection space. Thus each chain conformation is assigned an index corresponding to the group it belongs to. Such an index is called a cluster index (ID). We used density-based spatial clustering of applications with noise (DBSCAN)~\cite{DBSCAN} for each projection in four dimensional space of leading principal components (PCs). The cluster ID $n_i$ for a single chain at each time frame is always an integer, i.e., $n_i \in\{-1, 0, 1,2,\ldots, n_{cluster}-1\}$ for $i=1,2,\ldots, n_cM$, where the number of chain $n_c=2000$, the number of frames $M=150$ at each $T$, $n_{cluster}$ is a number of clusters found by DBSCAN ($\max(n_{cluster})=3$ in this work). $n_i=-1$ corresponds to the noise while $n_i\ge 0$ corresponds to the clusters found in the four-dimensional PCA projections using DBSCAN~\cite{DBSCAN}. The details of clustering, the rationale for choosing four dimensions in PC space,  the goodness of clustering are given in the SI (Sec.~S-V). DBSCAN determines the high-temperature states as sparse or ``noise'' (and assigns them with cluster ID =-1) and the low-temperature glassy state as a cluster(s) (Cluster IDs $\geq$ 0), see e.g. \Cref{fig:glass-transition-both}b-c. 
Then, we repeat this clustering on each chain present in the system (2000 chains) to confirm that the separation between liquid and  glassy state is consistent for all chains in the melt. 
To obtain a general estimate of the temperature at which this separation occurs, we calculate the average cluster ID $\langle n (T) \rangle$. At each temperature $T$, $\langle n (T) \rangle$ is given by $\langle n (T) \rangle =  \sum_{n_i=-1}^{n_{cluster}-1} n_i P(n_i, T)$, 
where $P(n_i,T)$ is the probability distribution of cluster IDs for all $n_c$ chains over $M$ frames at each $T$.  
In \Cref{fig:glass-transition-both}d $\langle n (T) \rangle$ shows a sharp transition around $T = 0.65\epsilon/k_B$.

The glass transition is often viewed as the process of falling out of equilibrium during cooling at a given rate or as the onset of ergodicity breaking. Above $T_g$ all states are accessible to the system, while below $T_g$ the system is arrested. Therefore, we expect the dissimilarity between low and high temperature regimes at or around $T_g$: giving rise to the sharp transition in the average cluster indices. Our result shows a signature of dynamic ergodicity breaking (Fig.~S1) indicated by a dramatic increase of equilibration time (see the VFT-plot in Ref.~\citenum{Hsu2019}) for each chain at the same temperature --- we report that as $T_g$. A similar signature of ergodicity breaking is reported recently using  Jensen--Shannon divergence metric for homopolymers~\cite{jin2022}. 

\begin{figure}
	\centering
        \centering
		\includegraphics[width=1.0\linewidth]{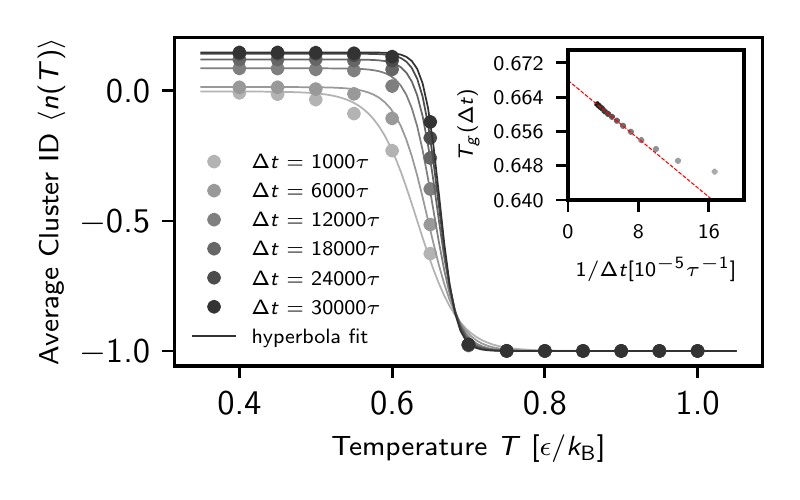}
		\caption{ DBSCAN for different selected observation time windows $\Delta t$, as indicated. The curves give the best hyperbola fit $g(T)$ going through the data.  
The inflection point of $g(T)$ shown in the inset gives the estimate of $T_g(\Delta t)$ at each $\Delta t$.
Extrapolating to $\Delta t \rightarrow \infty$, we obtain $T_g \approx 0.6680 \epsilon/k_B$.}
		\label{fig:pca_all_T}
\end{figure}

To extrapolate obtained results to  long time limits where the polymer chains are supposed to reach the diffusive regime, we repeated the analysis above for sixteen observation time windows $\Delta t$ ranging from $1000\tau$ to $3 \times 10^4\tau$. For each $\Delta t$, we have used $\Delta t/ t_{\rm lag}$ consecutive frames with $t_{\rm lag}= 200\tau$.
We see that the transition from liquid to glassy state becomes sharper with the increase of the observation time window. 
To quantify that we interpolate the data by a hyperbolic tangent function $g(T)=C(\Delta t)(1-\tanh(sT-d))/2-1$, where $s$ and $d$ are the fitting parameters, $C(\Delta t)$ is the gap between the two states at $T\gg T_g$ and $T \ll T_g$, respectively. The inflection point of $g(T)$ gives the estimate of $T_g(\Delta t)$ depending on  $\Delta t$. The behavior of average cluster ID vs. T as given  in \Cref{fig:pca_all_T} is similar to a typical behavior of magnetization vs. T for a finite-size 2D Ising model~\cite{landau2000importance}, and requires further investigation considering the existing discussion  in the literature~\cite{biroli2013perspective}. The finite-size (time) effect is often considered for analyzing data obtained from simulations of finite system sizes or limited computing times.
Taking into account this finite-time effect, we plot the estimates of $T_g(\Delta t)$ versus $1/\Delta t$ in the inset of \Cref{fig:pca_all_T}. We find a remarkable linear dependency, which allows for extrapolation to $\Delta t \rightarrow \infty$ and obtain $T_g \approx 0.6680 \epsilon/k_B$ as a best asymptotic estimate of $T_g$. This is in excellent agreement with the classical analysis of the temperature-dependent density (\Cref{fig:glass-transition-both}a). 

In order to interpret the obtained projections to the leading PCs, we calculate the correlation between the internal distances and the corresponding projection to PCs (see SI, Sec.~S-VII).  
The mostly-correlated distances vary with different chains, with no clear signature of any characteristic distance. Due to standardization of the distances (i.e., see $\mathbf{X}_c$ definition for details), PCA accounts for relative changes in the distances rather than the absolute displacement values. As a result, the projections to leading PCs are not dominated by only large distances. However, they are related to physically motivated measures such as $R_g$, $R_e$ (other physical properties can be also compared).

Note that we performed PCA on a single chain, followed by taking an average over all chains in the system. Performing PCA on $2000$ chains combined, we only observe the same Gaussian-like distribution within fluctuations, stemming from different chains, which is essentially independent of the temperatures (see SI, Fig. S7).

\begin{figure}
	\centering{
		\includegraphics[width=1.0\linewidth]{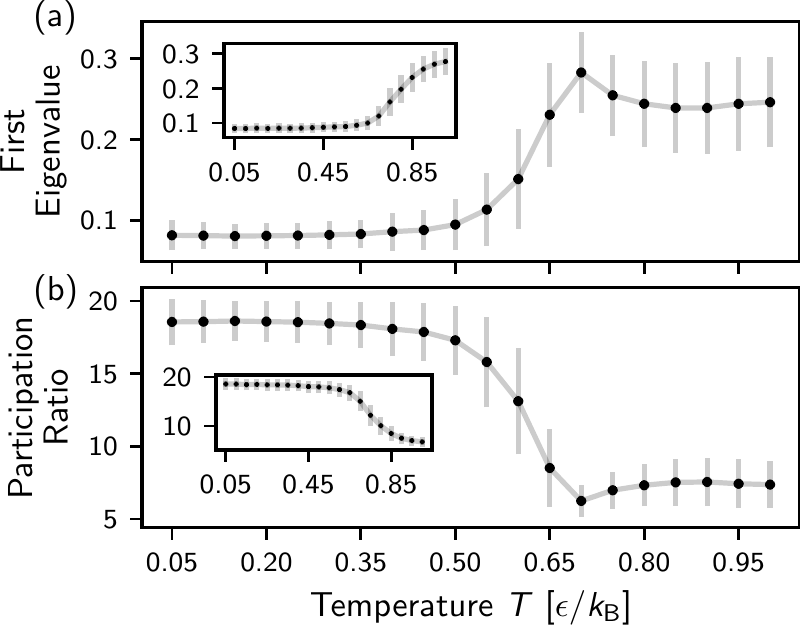}}
	\caption{ Analysis of each temperature independently. Mean values including deviations of the magnitude of first eigenvalues (a) and the participation ratio (b). Data taken from the time window between $200\tau$ and $3\times 10^4\tau$ (gray area in Fig. S1). The results for a shorter time upto $20 \tau$ (blue area in Fig. S1) are shown in the insets.}
	\label{fig:part_ratio_and_eigenvalues}
\end{figure}

\textit{Method II}. In the following, we change our approach and perform PCA for individual chains, but at different temperatures independently. In this way the new basis formed by eigenvectors $\mathbf{v}_{c,i}(T)$ differ for each temperature (see SI sec. Sec. S-IX, showing examples of first eigenvectors for Methods~I and II) and no information of individual chain conformations from other temperatures is accessible to Method~II. The resulting projections are shown in SI, Fig.~S9. Notably, for the majority of chains in the melt, we could observe the change from a completely random distribution of points in the projection to more ``clustered'' with the decrease of $T$. 
This behaviour can be quantified by the magnitude of the eigenvalues of PCA. In general, this magnitude is not a uniform value for independently projected data, but in our case all distances are standardised. Thus, we could average over the first eigenvalue for all projections (see \Cref{fig:part_ratio_and_eigenvalues}a), which shows a (weak) maximum close to $T_g$. This suggests that above $T_g$ large scale fluctuations dominate, while below $T_g$ fluctuations are dominated by many contributions from different, but short length scales.
As a more general criterion, we use the participation ratio (PR) defined at each temperature over 150 frames as $\rm{PR} = {(\sum_{i=1}^k \lambda_{c,i})^2}/{\sum_{i=1}^k \lambda_{c,i}^2}$, where $\lambda_{c,i}$ are eigenvalues sorted in the descending order (see \Cref{fig:part_ratio_and_eigenvalues}b). PR reflects decay rate of eigenvalues: the steeper is the change the smaller PR will be, if all $\lambda_{c,i}$ are equal then $PR = k$. A typical spectrum of eigenvalues $\lambda_{c,i}$ with different decay rates are plotted in SI Fig.~S4c.
The leading $k=25$ eigenvalues from $\textrm{min}(L,M)$ eigenvalues are counted to preserve at least $80\%$ data fluctuations in PCs. 
Results are averaged over all chains, deviations are shown as errorbars. 
 The increase in magnitude of the first eigenvalue (or the decrease in PR) on approaching $T_g$ can be related to an appearance of state separation in the system and change in a local structure as some recent studies suggest~\cite{godey2019local, jin2022}.
We argue that a prominent change in the monotonic behaviour of PR (or the first eigenvalue) is connected with a change in the nature of the fluctuations in the system: from local configurational rearrangements \change{(the rearrangement of parts of chain conformations)} above $T_g$ to only localized fluctuations along the chain below $T_g$ (similar to observations in metallic glasses \cite{Smith2017}). As a result, more dimensions are needed to describe the random motion below $T_g$. To test the hypothesis about local structural changes above $T_g$, we perform the same analysis on simulation trajectories within a relatively short time window between $0.2\tau$ and $20\tau$ (blue area in Fig. S1).
Results are shown in the inset in \Cref{fig:part_ratio_and_eigenvalues}, respectively. We no longer see the non-monotonic signature around $T_g$ since chains remain in their initial conformations within $1\sigma$ fluctuation in such a small time window.
Projections of short-time data from individual temperatures are given in SI, Fig. S10. 

In general, with method II, one can perform PCA on simulation trajectories at each temperature and monitor the eigenvalues and PR. Once we observe the non-monotonic change in both quantities around $T_g$, further simulations at lower temperatures are not required to localize $T_g$. 

In summary, we propose a new approach for determining the glass transition temperature from molecular dynamics simulation data with a fixed stepwise cooling protocol. The proposed data-driven protocol requires minimum input parameters and defines $T_g$ in a robust and transferable fashion. Our analysis focuses on the information about structural fluctuations  at the level of individual chains to identify the glass transition temperature and predict $ T_g$ for infinite simulation time from moderate simulation trajectories. We hypothesize that the relative distance fluctuations measured by the PCA may be directly correlated with the configurational entropy in the space of a single chain~\cite{Takano2017}. The method can be applied to a wide range of systems with microscopic/atomistic information. The generality of our approach could be tested with different dimensionality reduction and clustering methods. Further work in this direction is in progress.

\textbf{Acknowledgement}\\
We acknowledge open-source packages Numpy~\cite{Harris2020},
Matplotlib~\cite{Hunter2007}, Scikit-learn~\cite{scikit-learn} used in this work. The authors thank Michael A. Webb, Saikat Chakraborty and Daniele Coslovich for insightful discussion. 
The authors also thank Aysenur Iscen and Denis Andrienko for critical reading of the manuscript. 

	\textbf{Supporting information}\\
Supporting Information includes: simulation details (S-I), static properties of polymer melts (S-II), additional information on the data-driven approach (S-III), variance explained ratio of pca projections (S-IV), detailed description of clustering (S-V), principal component analysis on combined chains (S-VI), interpretation of leading principal components (S-VII), projections of a chain after performing PCA independently at each temperature (S-VIII), results with reduced number of descriptors (S-IX).



%

\clearpage
\pagebreak
\onecolumngrid








\newcommand{\hly}[1]{\colorbox{yellow}{#1}}
\newcommand{\atreyee}[1]{\textcolor{magenta}{#1 (atreyee)}}
\newcommand{\changes}[1]{#1}

\begin{center}
\textbf{\large Supporting Information to ``Data-driven identification and analysis of  
the glass transition in polymer melts''}
\vskip 0.5truecm
Atreyee Banerjee, Hsiao-Ping Hsu, Kurt Kremer, and Oleksandra Kukharenko
\vskip 0.1truecm
{\it Max Planck Institute for Polymer Research, Ackermannweg 10, 55128 Mainz, Germany}
\end{center}








\renewcommand{\thetable}{S\arabic{table}}
\renewcommand{\thefigure}{S\arabic{figure}}
\setcounter{equation}{0}
\setcounter{figure}{0}
\setcounter{table}{0}
\setcounter{page}{1}
\makeatletter
\renewcommand{\theequation}{S\arabic{equation}}
\renewcommand{\thefigure}{S\arabic{figure}}


\section{Simulation Details}
\label{sec:si:simulation}
\begin{figure}[h!]
	\centering
	\includegraphics[width=0.5\linewidth]{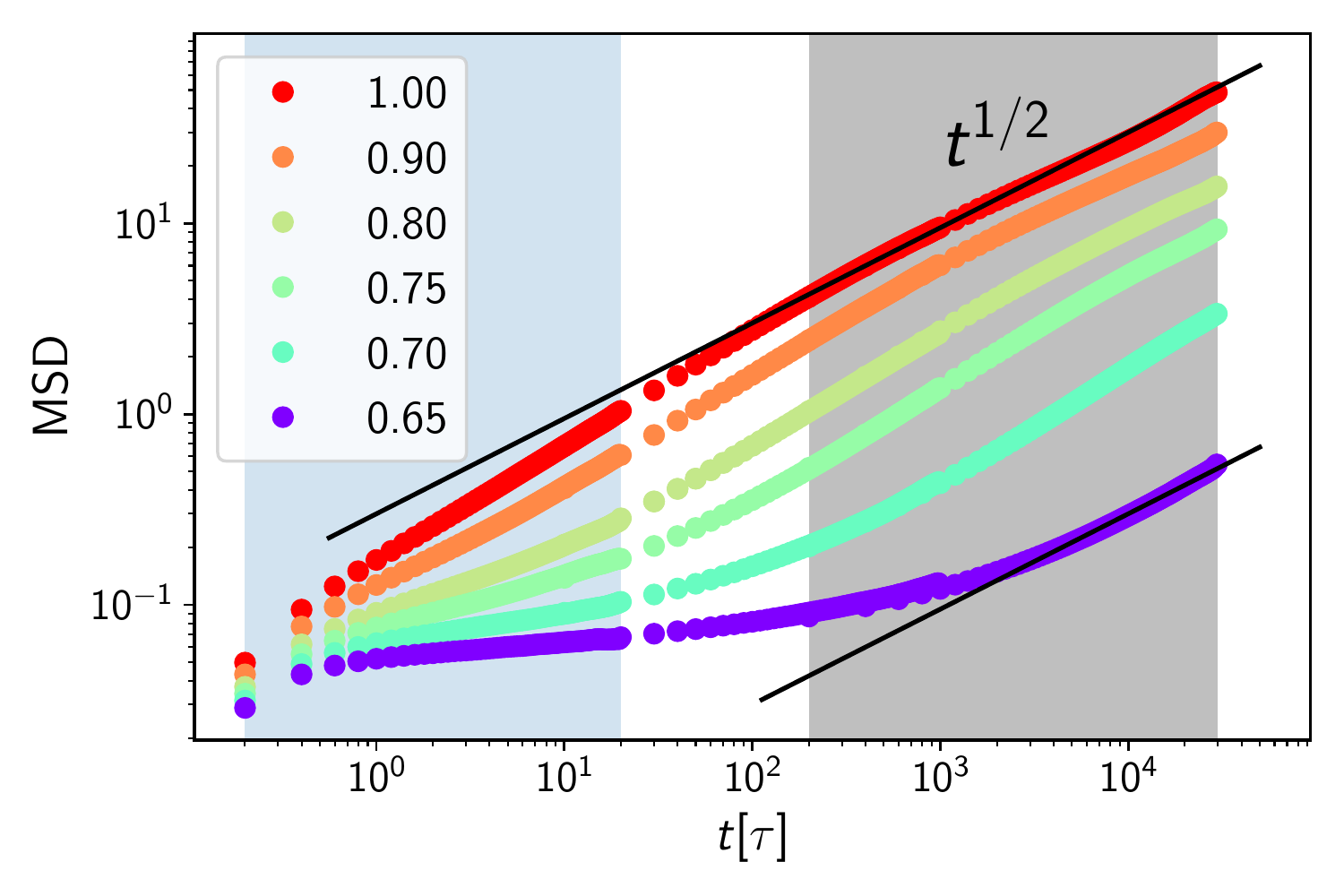}
	\singlespace
	\caption{Adapted from \cite{Hsu2019}: Mean square displacements (MSD) $g_1(t)$ of inner $12$ monomers as a function of time for bulk polymer melts containing $n_c=2000$ chains of length $n_m=50$ monomers at various selected temperatures $T[\epsilon/k_B]$, as indicated. For testing our data-driven approach, we take MD simulation trajectories at every $200\tau$ in the time frame between $200\tau$ and $3\times 10^4\tau$ (shaded gray colored region). We have also taken the trajectories at every $0.2\tau$ in the time frame between $0.2\tau$ and $20\tau$ (shaded blue colored region) for comparison. The Rouse-like scaling law $g_1(t) \sim t^{1/2}$ is represented by straight lines.}
	\label{fig:MSD_timescale}
\end{figure}

For studying polymer melts of semiflexible chains in bulk, in confinement and with free surface,
a new coarse-grained model (a new variant of bead-spring model) was developed recently~\cite{Hsu2019}. A short-range attractive potential between non-bonded monomer pairs is added such that the pressure $P$ can be tuned at zero. For keeping the chain conformations which only weakly depend on the temperature $T$, the conventional bond-bending potential~\cite{Faller1999,Everaers2004} is replaced by a new bond-bending potential, paramterized to conserve conformational properties at $T=1 \epsilon/k_B$. 
This model was tested by studying a bulk polymer melt of weakly semiflexible chains under cooling.
For such a system, the monomer density at pressure $P=0.0 \epsilon/\sigma^{-3}$ is $\rho=0.85 \sigma^{-3}$ and the entanglement length $N_e=28$ monomers.  Starting from a fully equilibrated polymer melt consisting of $n_c=2000$ polymer chains of $n_m=50$ monomers with the Kuhn length $\ell_K \approx 2.66\sigma$~\cite{Hsu2016}, molecular dynamics (MD) simulations were performed in the NPT ensemble at $P\approx 0\epsilon/\sigma^3$ and constant temperature $T$ following a stepwise cooling strategy. The temperature was reduced in steps of $\Delta T=0.05\epsilon/k_B$ from $T=1.0\epsilon/k_B$ to $0.05\epsilon/k_B$ with a relaxation time between each step of $\Delta t=60000\tau \approx 8.3 \tau_R$ (chain conformations were stored every $500\tau$), $\changes{\tau_R=\tau_0n_m^2} \approx 7225\tau$ being the Rouse time of relaxing the overall chains at $T= 1 \epsilon/k_B$. \changes{The characteristic relaxation $\tau_0$ is determined from the mean square displacement (MSD) of inner monomers~\cite{Hsu2016}.} This resulted in a cooling rate of $\Gamma=\Delta T/\Delta t = 8.3 \times 10^{-7} \epsilon/(k_B\tau)$. Here $k_B$ is the Boltzmann factor, $\sigma$, $\epsilon$ and $\tau=\sigma\sqrt{m/\epsilon}$ with a monomer mass of $m=1$ are the Lennard-Jones units of length, energy, and time, respectively. 
Starting from the last configuration obtained from the NPT run at each $T$, further MD simulations \changes{choosing the time step $\delta t=0.01\tau$} were performed in the NVT ensemble. Estimates of MSD of inner $12$ monomers, $g_1(t)$, characterizing the mobility of chains are shown in \Cref{fig:MSD_timescale}. All MD simulations are performed using the package ESPResSo++~\cite{Espressopp,Espressopp20}.

\section{Static Properties of Polymer Melts}
\label{si:sec:re_rg}

\begin{figure}[!h]
	\centering
		\subfigure[]{
			\includegraphics[width=0.42\linewidth]{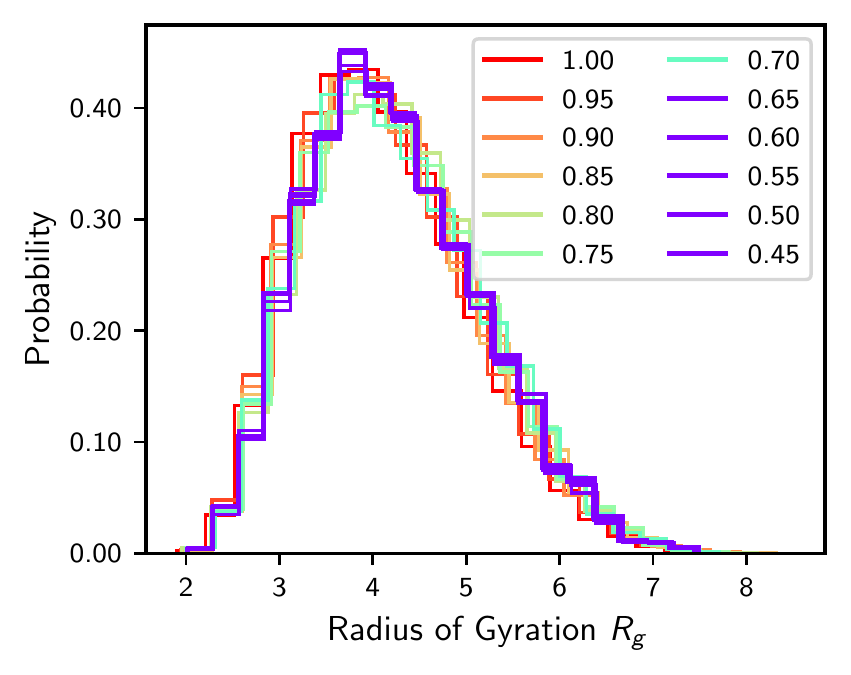}}\hspace{0.5truecm}
		\subfigure[]{
		\includegraphics[width=0.42\linewidth]{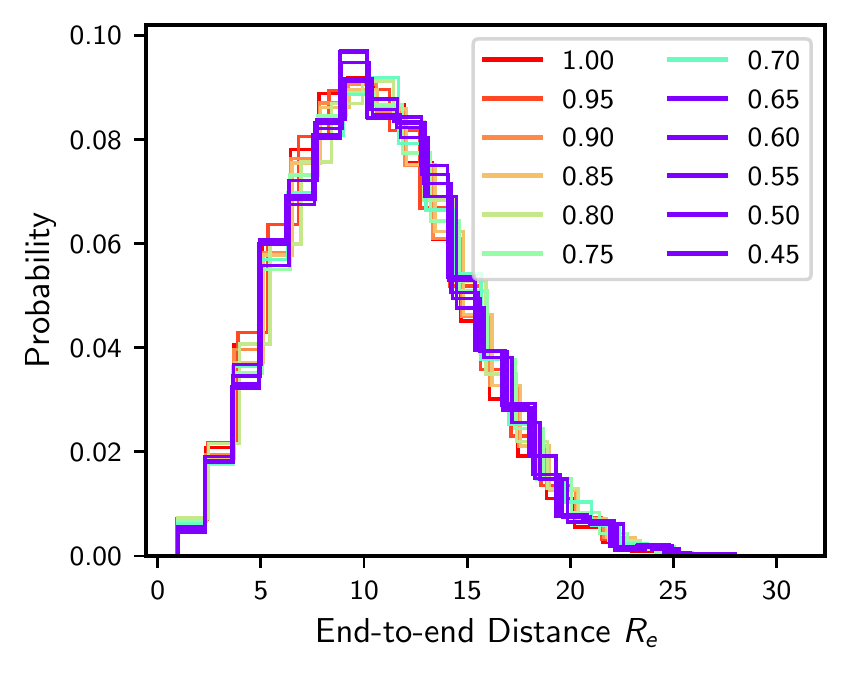}}
	\caption{Probability distributions of radius of gyration (a) and end-to-end distance (b) at several selected temperatures $T$, as indicated.}
	\label{fig:Rg_dis}
\end{figure}

The conformation of a polymer chain usually is described by the radius of gyration $R_g$ and end-to-end distance $R_e$ as follows,
\begin{equation}
\label{si:eq:r_g}
    R_g = \sqrt {\frac{1}{n_m}\sum_{i=1}^{n_m}({\bf r}_{i}-{\bf r}_{\rm{c.m}})^2} \quad \textrm{with} 
    \quad 
    {\bf r}_{\rm{c.m.}} = \frac{1}{n_m}  \sum_{i=1}^{n_m} {\bf r}_{i}
\end{equation}
and
\begin{equation}
\label{si:eq:r_e}
    R_e = \sqrt{ ({\bf r}_{n_m}-{\bf r}_{1})^2}
\end{equation}
where ${\bf r}_{i}$ is the coordinate of $i$th monomer in the chain, and ${\bf r}_{\rm{c.m.}}$ is the center of mass of chain. As mentioned above, the profiles of probability distributions of $R_g$ and $R_e$ for all $2000$ chains in the melt should remain the same within fluctuation at all temperatures, as shown in \Cref{fig:Rg_dis}. 
\vspace{2.0truecm}
\newpage
\section{Data-driven approach}
\label{sec:si:data-driven}

\begin{figure*}[h!]
	\centering
    \includegraphics[width=1.0\linewidth]{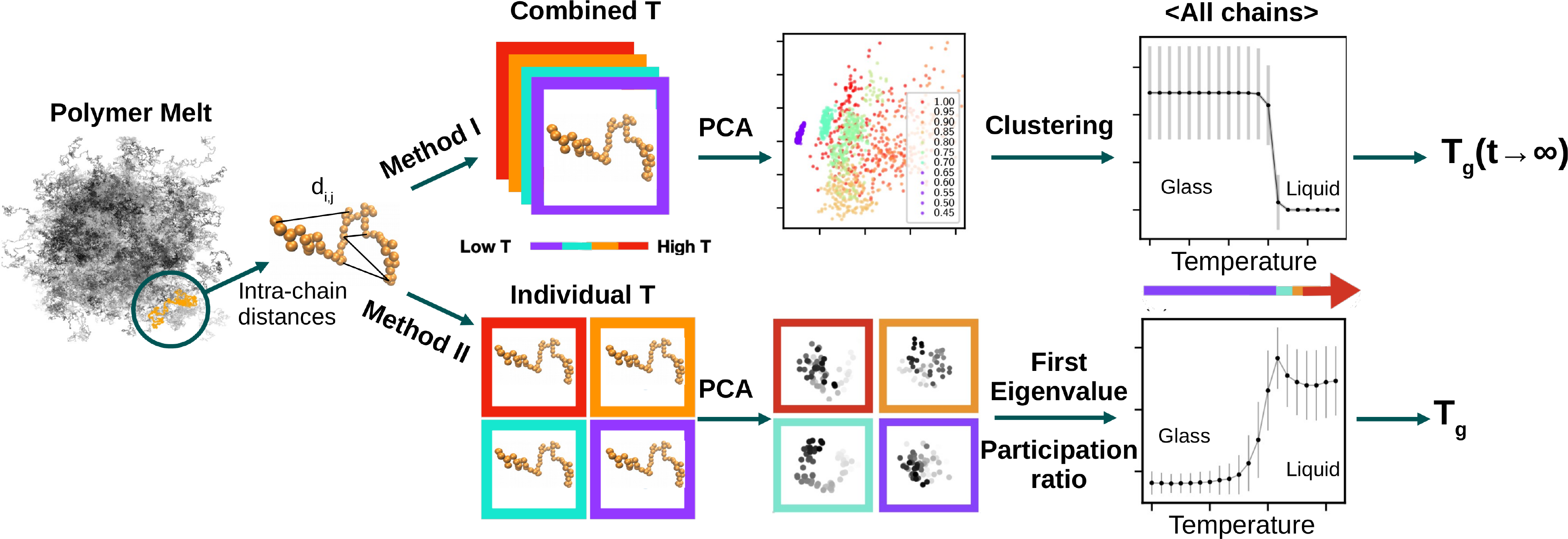}
	\singlespace\caption{Schematic representation of the two methods employed in the paper for the determination of glass transition temperature $T_g$. The workflows uses intra-chain distances of individual chains from the melt. It can be applied in two independent ways: by projecting with PCA combined data from temperatures followed by clustering, where the change in cluster indexes indicates $T_g$ (Method I, upper raw). Or by applying PCA to each temperature separately and using changes in leading eigenvalues or participation ratio as the criteria to define $T_g$ (Method II, lower raw).}
 \label{workflow}

\end{figure*}

\vspace{1.5truecm}
\section{Variance Explained Ratio of PCA projections}
\begin{figure*}[!h]
	\centering
 	\subfigure[]{
\includegraphics[width=0.33\linewidth]{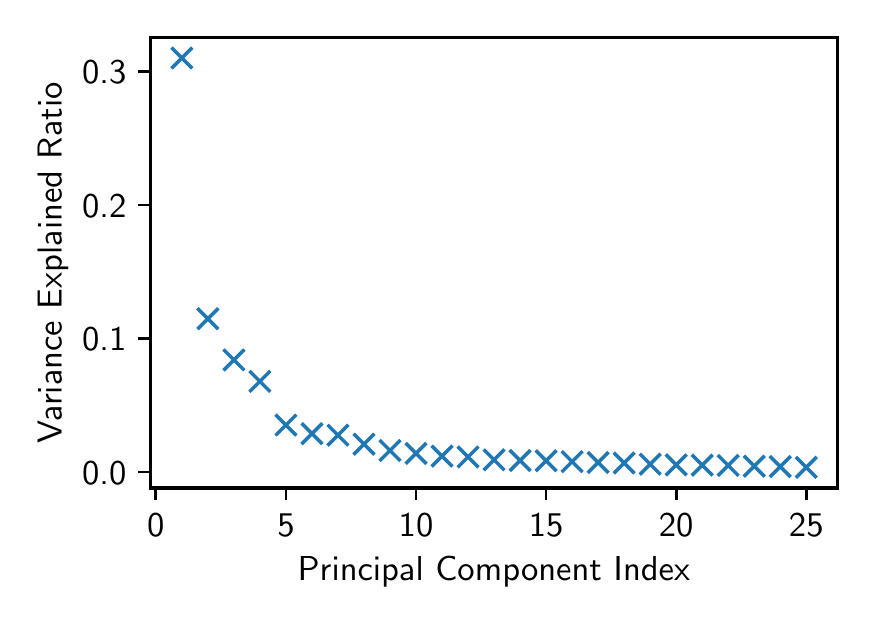}}
 	\subfigure[]{
\includegraphics[width=0.33\linewidth]{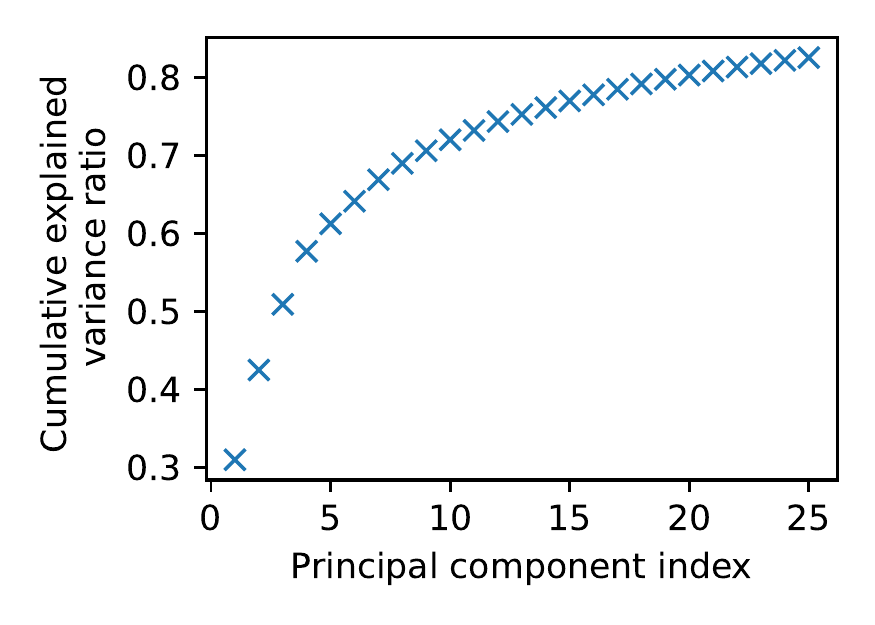}}
\subfigure[]{
\includegraphics[width=0.3\linewidth]{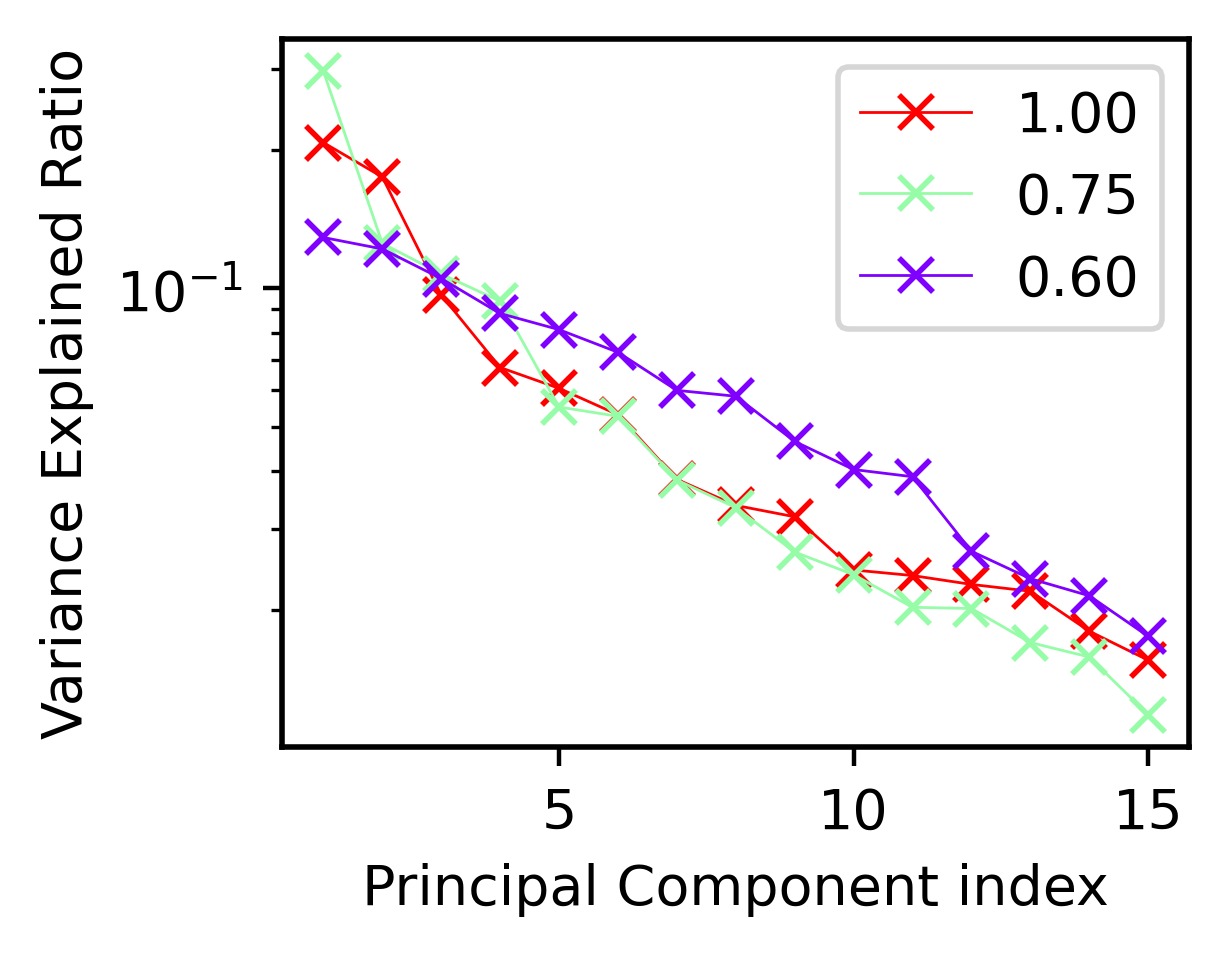}}
\singlespace\caption{\changes{ (a) \change{Variance explained ratio} plotted vs. principal component index for data combined from all $T$ (Method I).  Principal components are ranked by the amount of variance they capture in the original data. 
The gap before the fifth \change{component}, suggesting the leading 4 components are important.  (b) Cumulative summation of the explained variance (presented in (a)) shows that first four PCs describe $\approx60\%$ of variance in the data. (c) \change{ Variance explained ratio} plotted vs. principal component index for three selected temperatures $T[\epsilon/k_B] = 1.0, 0.75, 0.6$ (Method II). The decay rates of eigenvalues are different. The participation ratio reflects this decay being highest when the decay rate is slowest.}}
\label{fig:explained_varience}
\end{figure*}

Four leading PCs containing $\approx 60\%$ of a variance of the data based on the gap in the variance explained ratio\footnote{The ratio between the variance of each principal component and the total variance.} (\Cref{fig:explained_varience}). 

 \changes{\section{Clustering method -- DBSCAN}}
 \label{sec:si:clustering_method}
 \changes{The projected data is grouped using the unsupervised learning technique known as clustering analysis. Typically grouping is done based on some common property, i.e. closeness in space, similarities in densities (same number of neighbours within a cut-off separated by low-density regions), etc. The choice of clustering technique is then motivated by the properties based on which the data should be grouped, properties of the data itself (e.g. the number of data points, dimensionality etc.), as well as by some additional knowledge about the system. All clustering techniques require initial parameters to define clusters.  Then they try to arrange the data points into groups. As a result each data point is assigned an index corresponding to the group it belongs to. Such an index is also called a cluster index (ID).} 

 \changes{In our case, each data point in the projection corresponds to one configuration of a chain at a given time and temperature. Expecting that the configurations at lower temperatures will be much closer together and disordered at higher temperatures for clustering we chose the density-based spatial clustering of applications with noise (DBSCAN)\cite{DBSCAN}. The acronym was given by the authors of the clustering algorithms (Martin Ester, Hans-Peter Kriegel, J\"{o}rg Sander and Xiaowei Xu, 1996) and for our case, it refers to distances and densities in four-dimensional projected PCA space, where now one chain is represented as a point in four dimensions. It is designed to find high dense regions in space as separate clusters and assign all sparse points as unclustered or ''noise''.}\\

 \textbf{DBSCAN parameters.} DBSCAN groups points that are close to each other based on a distance measure (usually Euclidean distance\footnote{ For the points $p$ and $q$ with Cartesian coordinates $p_i$ and $q_i$ in $n-$dimensional Euclidean space, the distance is defined as 
 \begin{equation*}
     d(p,q)={\sqrt {(p_{1}-q_{1})^{2}+(p_{2}-q_{2})^{2}+\cdots +(p_{n}-q_{n})^{2}}}.
 \end{equation*}}) and a minimum number of points within some cut-off.

There are two parameters one needs to specify for DBSCAN:
\begin{enumerate}[1)]
    \item A cut-off value to define proximity in space -- minimum distance ($\epsilon_d$): indicates the radius  within which points to be regarded as neighbors. Two points are regarded as neighbors if the distance between them is less than or equal to this cut-off value $\epsilon_d$. Small $\epsilon_d$ is used for defining denser clusters. On the other hand, if $\epsilon_d$ is too high, the majority of the data will be in the same cluster. There are strategies to choose $\epsilon_d$~\cite{Schubert2017}.
    \item Number of nearest neighbors (NN): number of points within distance $\epsilon_d$ to make the smallest cluster. As a rule of thumb, the number of minimum points should be greater than a dimensionality of points and typically chosen as twice the number of dimensions, but it may be necessary to choose larger values for different data sets.
\end{enumerate}


DBSCAN has a number of advantages: it can find non-linearly separable clusters (clusters of any shapes); the number of clusters is not a prior parameter in the method (compared to many other methods); and most importantly for us, it has the notion of noise.\\

  \changes{\textbf{Evaluation of clustering results.} To quantify the quality of the clustering results depending on the chosen parameters we used 
V-measure (or normalised mutual information) score~\cite{rosenberg2007}. For this measure, reference cluster IDs for each point are provided. Those reference IDs are called true labels or ground truth. For our system, we define reference states (ground truth \change{from Ref.~\cite{Hsu2019} using the volume change}) as cluster IDs = -1 for the states above $T_g$ ($T > 0.65\epsilon/k_B$), otherwise  cluster IDs = 0.}

 \changes{The V-measure is an average of the other two measures: homogeneity and completeness. Homogeneity 
is defined using Shannon’s entropy.
\begin{equation}
    hom = \begin{cases}
      1 & \text{if $H(C) = 0$ }\\
      1 - \frac{H(C|K)}{H(C)}, & \text{otherwise.}
    \end{cases}
\end{equation}
where $C$ is the target clustering (ground truth), $H(C)=  -\sum _{c=1}^{|C|} \frac{n_{c}}{M} \log \frac{n_{c}}{M}$, $H(C|K)=  -\sum _{c=1}^{|C|} \sum _{k=1}^{|K|} \frac{n_{ck}}{M} \log \frac{n_{ck}}{n_k}$, $M$ is the size of a data set,  $n_{ck}$ is number of samples with the cluster ID $c$ in cluster $k$ and $n_k$ the total number of samples in cluster $k$. The homogeneity is equal to 1 when every sample in cluster $k$ has the same cluster ID $c$. $0 \leq hom \leq 1$, with low values indicating a low homogeneity.}

 \changes{
Completeness measures whether all similar points are assigned to the same cluster, it is given by
\begin{equation}
    comp = \begin{cases}
      1 & \text{if $H(K) = 0$ }\\
      1 - \frac{H(K|C)}{H(K)}, & \text{otherwise.}
    \end{cases}
\end{equation}
The completeness is equal to  1 when all samples of cluster ID $c$ have been assigned to the same cluster $k$, $0 \leq comp \leq 1$.}
 \changes{
Normalised mutual information or V-measure is a measure of the goodness of a clustering algorithm considering the harmonic average between homogeneity and completeness. It is given by
\begin{equation}
    {\rm V_{measure}} = 2*\frac{hom*comp}{hom+comp}.
\end{equation}
$ 0 \leq \rm {V_{measure}} \leq 1$, where high values indicate good clustering.}

\begin{figure}[h!]
        \centering
                \includegraphics[width=0.8\linewidth]{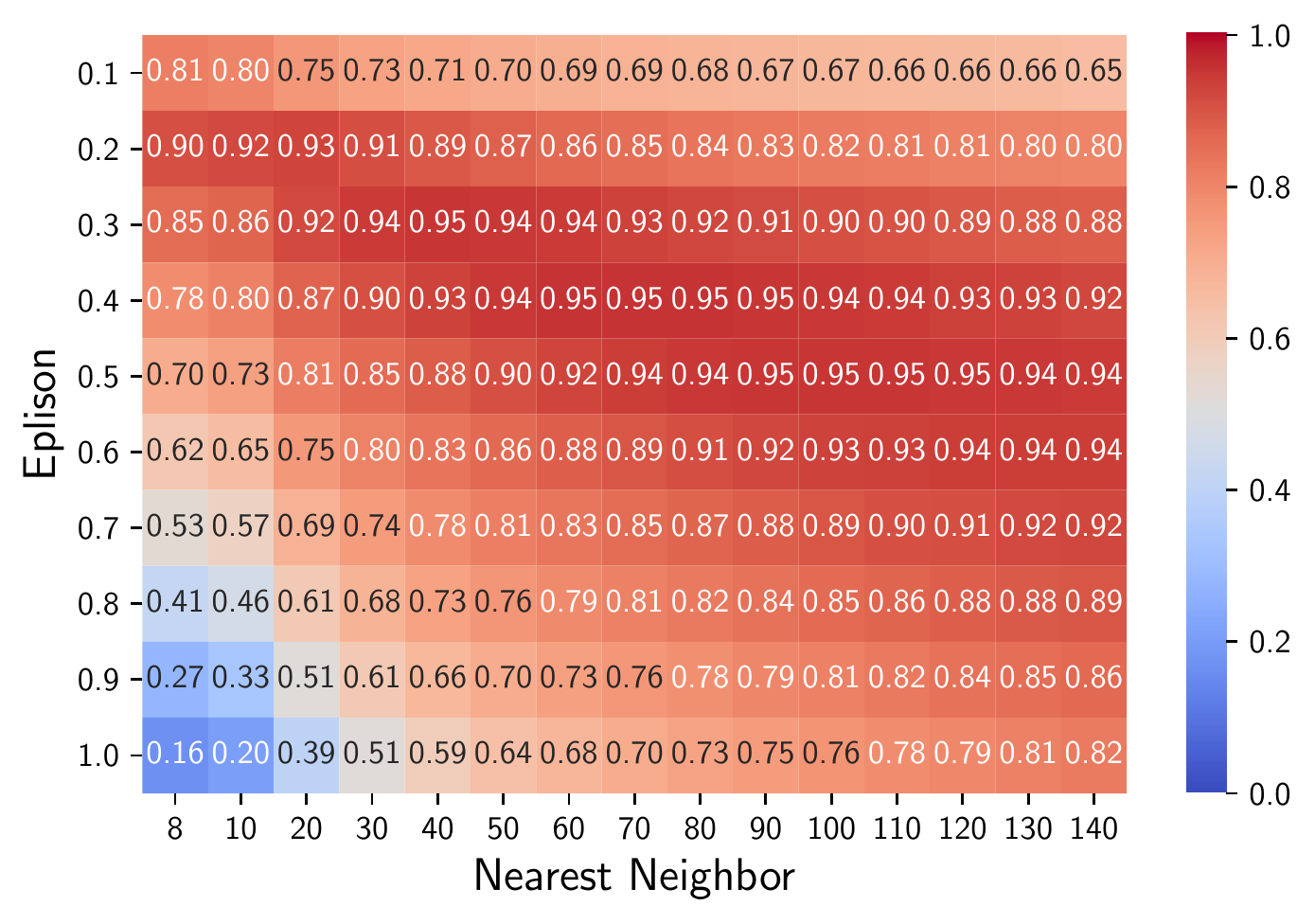} \label{fig:V-measure-pca}
\singlespace\caption{ Heat-map of V-measure on the PCA embedding averaged over all chains \change{with Method~I, i.e. combined temperatures using $3000$ frames}. 
The best parameter set for DBSCAN is chosen where the V-measure value is maximum.}
\label{fig:V-measure}
\end{figure}

 \changes{
The clustering score was computed using the python scikit-learn package~\cite{scikit-learn}.
As shown in \Cref{fig:V-measure}, 
it is possible to 
obtain good clustering scores for quite a big range of NN and $\epsilon_d$ values on the PCA embedding. For the data shown in the main text we used the Euclidean distance ($\epsilon_d=0.3$) to create a neighbourhood and the minimum number of points ($NN =40$) to form a dense region.} 

\begin{figure*}[t!]
	\centering
	\subfigure[]{
\includegraphics[width=0.45\linewidth]{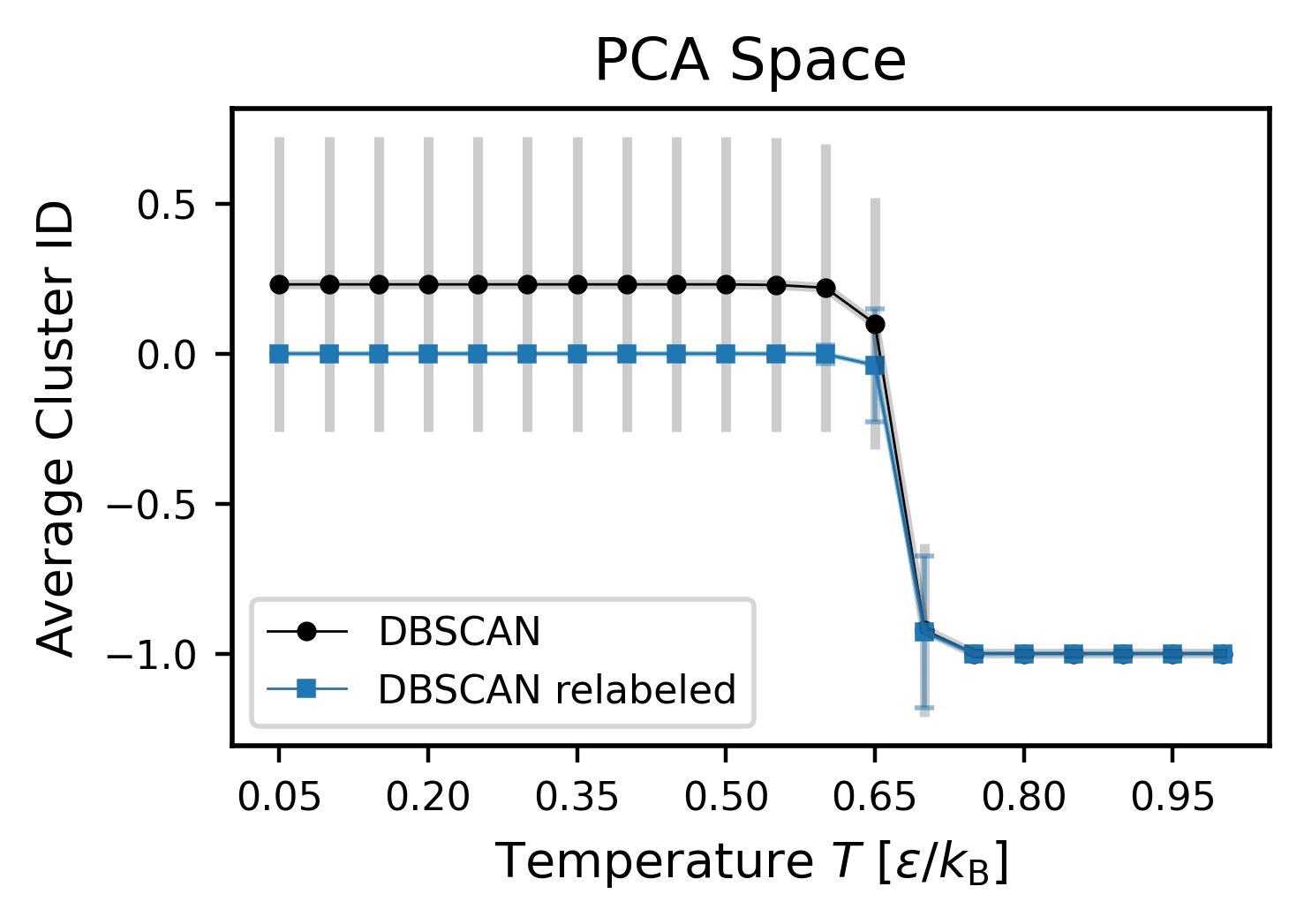}}
	\subfigure[]{
\includegraphics[width=0.47\linewidth]{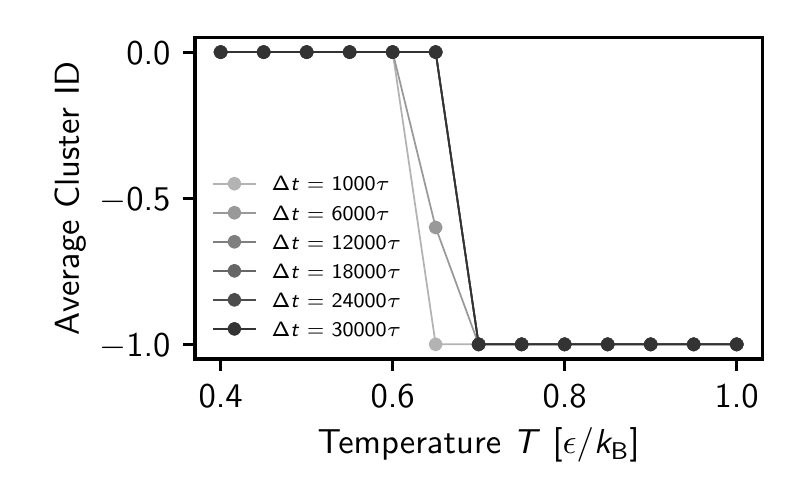}}
\singlespace \caption{Average clustering results in PCA space. \changes{ (a) Average clustering results by defining the error bar as deviation from the average cluster ID (black). If we relabel the clusters (blue) to make sure that there is only one cluster in the glassy state, we immediately observe that the error bars (blue bars) in the glassy state drop down to zero and they are maximum near the transition.} (b) Averaging using median instead of mean for DBSCAN cluster indices (IDs). The transition at $T_g$ becomes even sharper with this type of averaging.}
\label{fig:errorbars-pca}
\end{figure*}

\vspace{1.0truecm}
 \changes{
\textbf{Defining average cluster ID $\langle n (T) \rangle$ }. We perform DBSCAN clustering on the 4-dimensional projections to PCA space for each chain separately. DBSCAN defines the high-temperature states as sparse or ''noise'' (and assigns them with cluster ID = -1) and the low-temperature glassy state as a cluster(s) (Cluster IDs $\geq 0$). The change in cluster indices after $T = 0.65\epsilon/k_B$ is prominent (Figure~1d). Then we repeat this clustering on each chain present in the system (2000 chains), meaning each chain at each frame and temperature will get a cluster ID value (in our example for PCA space cluster IDs are -1, 0, 1 or 2). 
We observe that for some of the projections in the glassy state, there is more than one group/cluster found corresponding to cluster IDs 0, 1, 2.
Then we average over all cluster IDs for all chains  $\langle n (T) \rangle$ that were simulated at each temperature. If we use the mean value for averaging, the average cluster IDs are higher than zero and reflect that more than one cluster was found for this temperature. Such deviation in cluster IDs is shown in large constant error bars in \Cref{fig:errorbars-pca}a (black line).  
If we assume that there can be only one cluster in a glassy state and combine all the cluster IDs with 0, 1 or 2, the error bars in the glassy state drop down to zero and they are maximum near the transition region, where assigning the points to a group is the most challenging (\Cref{fig:errorbars-pca}a (blue line)). 
If we use the median for averaging we do not have such deviations from zero (\Cref{fig:errorbars-pca}b).}

\section{PCA on combined chains}

\begin{figure}[h!]
	\centering
	\subfigure{
		\includegraphics[width=0.50\linewidth]{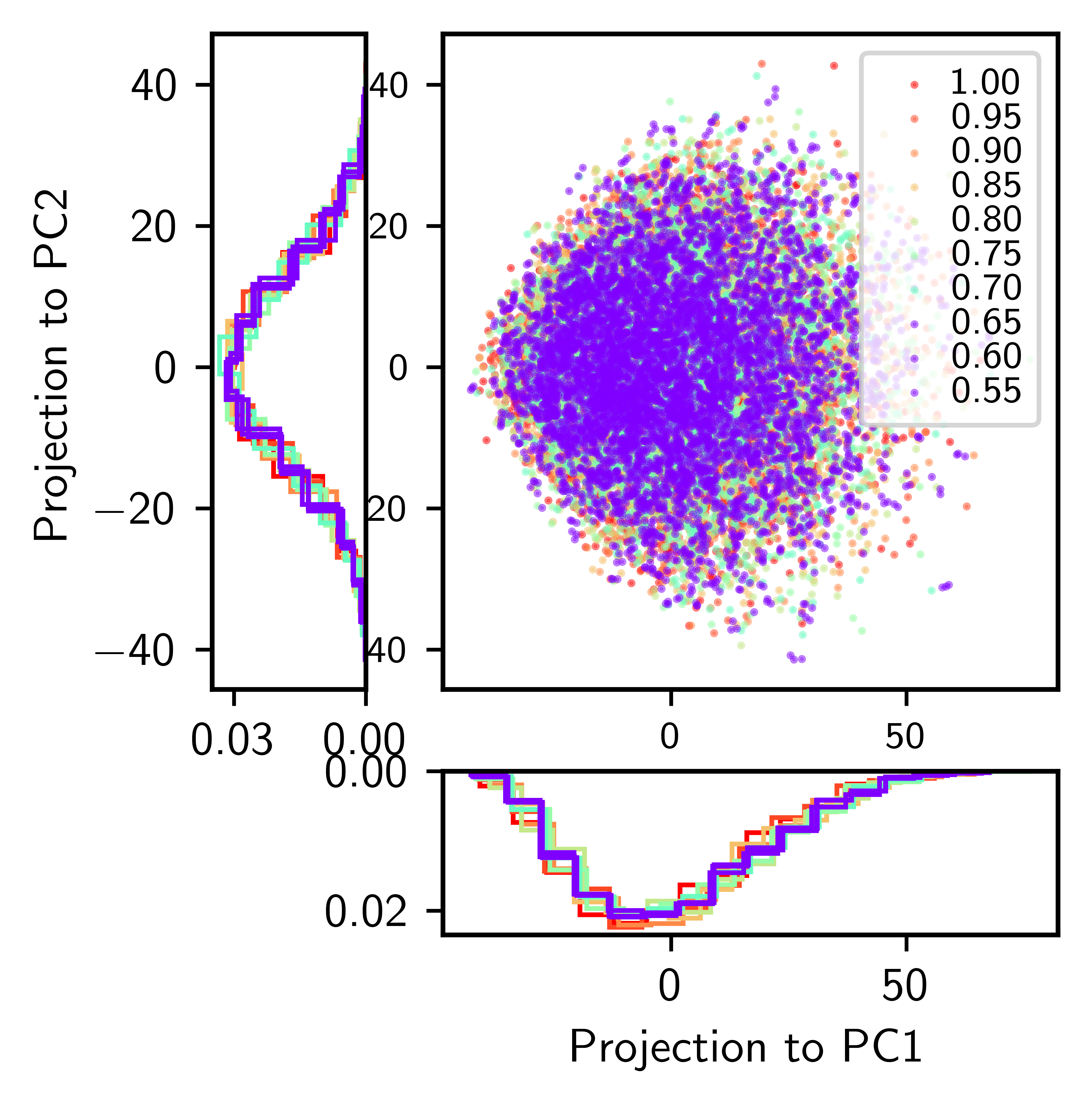}}
	\singlespace\caption{PCA projections of all chains, and the distributions of PCs for several selected $T$, as indicated. No temperature-dependent separation in the low dimensional projections is observed.}
	
	\label{fig:all_chain_together}
\end{figure}

In the main text, we perform PCA on intra-chain distances for an individual chain over simulation frames, followed by taking an averaging over all chains presented in the system. Here we perform PCA on all $2000$ chains together. We use the input data matrix $\mathbf{X} \in \mathbb{R}^{M \times L}$, where $L$ is the number of descriptors (e.g. internal distance of a single chain of chain length $n_m=50$: $L = n_m \times(n_m-1)/2$ = 1225, same as before), and $M$ is the number of observations (e.g. number of chains multiplied by number of temperatures, $2000 \times 20=40000$).
We observe the Gaussian-like distribution for all temperatures in \Cref{fig:all_chain_together} within fluctuations. 


\section{Interpretation of the projection to Leading Principal Components}

\begin{figure}[!h]
	\centering
 \subfigure[]{
\includegraphics[width=0.4\linewidth]{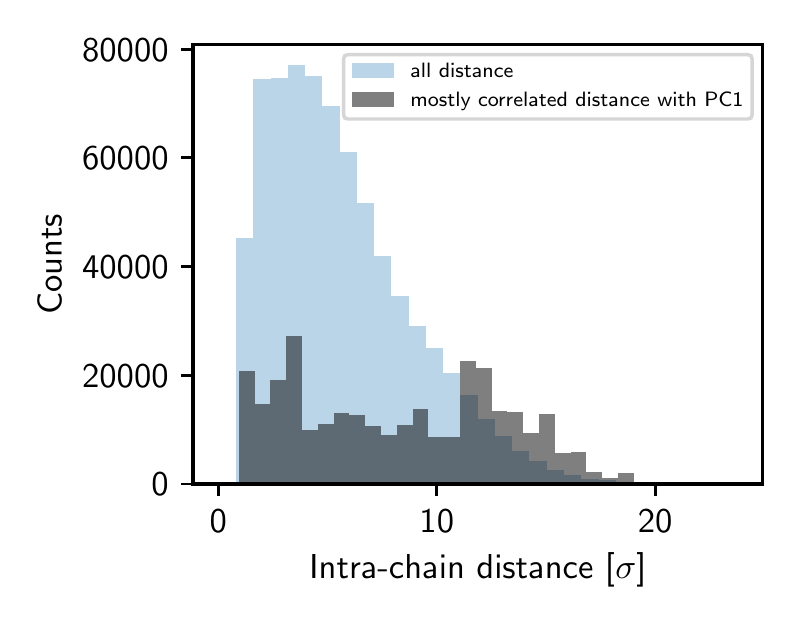}}
 \subfigure[]{
\includegraphics[width=0.40\linewidth]{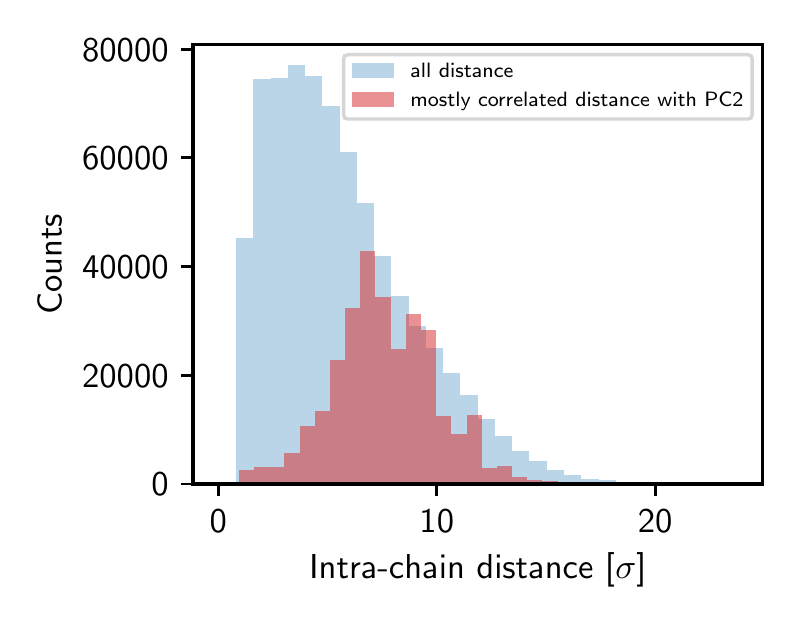}}
\subfigure[]{
\includegraphics[width=0.42\linewidth]{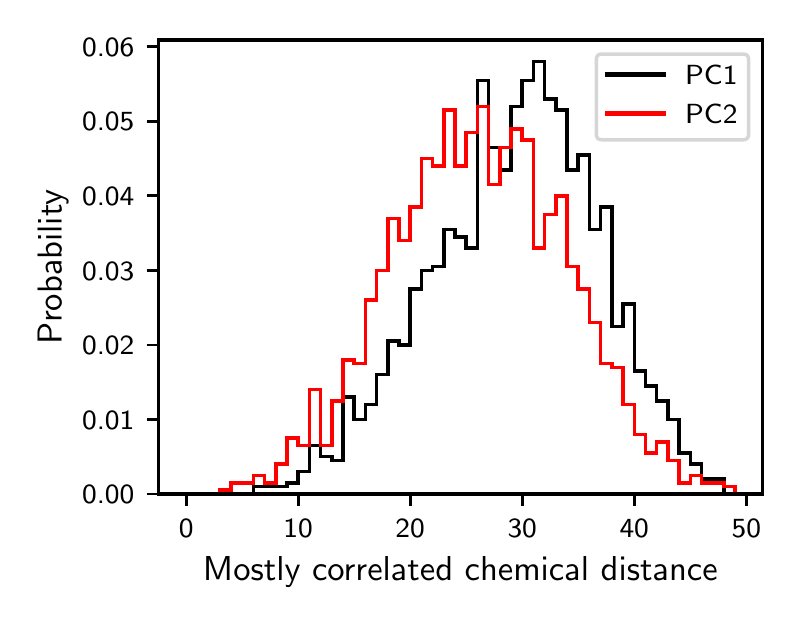}}
\subfigure[]{
\includegraphics[width=0.45\linewidth]{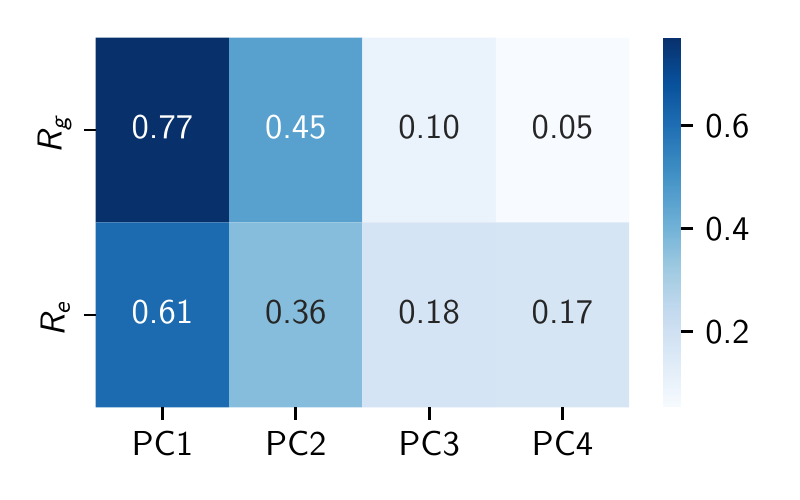}}

\singlespace\caption{ \changes{Distributions of intra-chain distances mostly correlated (correlation higher than 0.9) with projections to the first (a) and second (b) principal components for all chains compared with all intra-chain distances distributions (blue bars).} (c) Distribution of respective chemical distances of intra-chain distances highly correlated with the projection of two leading PCs. No preferences for longer/shorter distances ranges can be observed. d) Pearson correlation coefficients between projections to the leading PCs, a radius of gyration ($R_g$), and an end-to-end distance ($R_e$).}
\label{fig:PC_hig_corr}
\end{figure}

 \changes{
In order to interpret the obtained projections to PCs, we calculate the Pearson correlation coefficient~\cite{Pearson1895} between the input features (\change{intra-chain distances in space}) and the projections to leading PCs (first and second) \change{obtained by Method I}. 
The most correlated intra-chain distances and their locations in the chain \change{(chemical distance along the chain)} are identified for all chains and 
their probability distribution functions for all chains are shown in \Cref{fig:PC_hig_corr}a-c. 
There are no characteristic positions in a chain (e.g. end monomers) or distances which highly correlate with the projections to leading PCs. For most of the chains, the chemical distance with the highest correlations is normally distributed with a peak of around 30 monomers. Moreover, some intermediate distances, but not the longest, are dominant. This can be also seen in \Cref{fig:PC_hig_corr}a,b, where intra-chain distance distributions have no peaks at higher values. We should stress once more that due to standardisation of the distances PCA accounts for relative changes rather than the absolute displacement values which suggests that the rearrangements in long, medium, and short ranges have equal importance.}

\Cref{fig:PC_hig_corr}d shows averaged Pearson correlation coefficients between the projections to the leading PCs, a radius of gyration $R_g$ (\Cref{si:eq:r_g}), and an end-to-end distance $R_e$ (\Cref{si:eq:r_e}). Results are obtained by taking the average over all 2000 chains. We find that both $R_g$ and $R_e$ are correlated with the first PC.

\clearpage
\section{Projections of A Chain after Performing PCA independently at each temperature}

\begin{figure}[!htbp]
	\centering
		\includegraphics[width=0.62\linewidth]{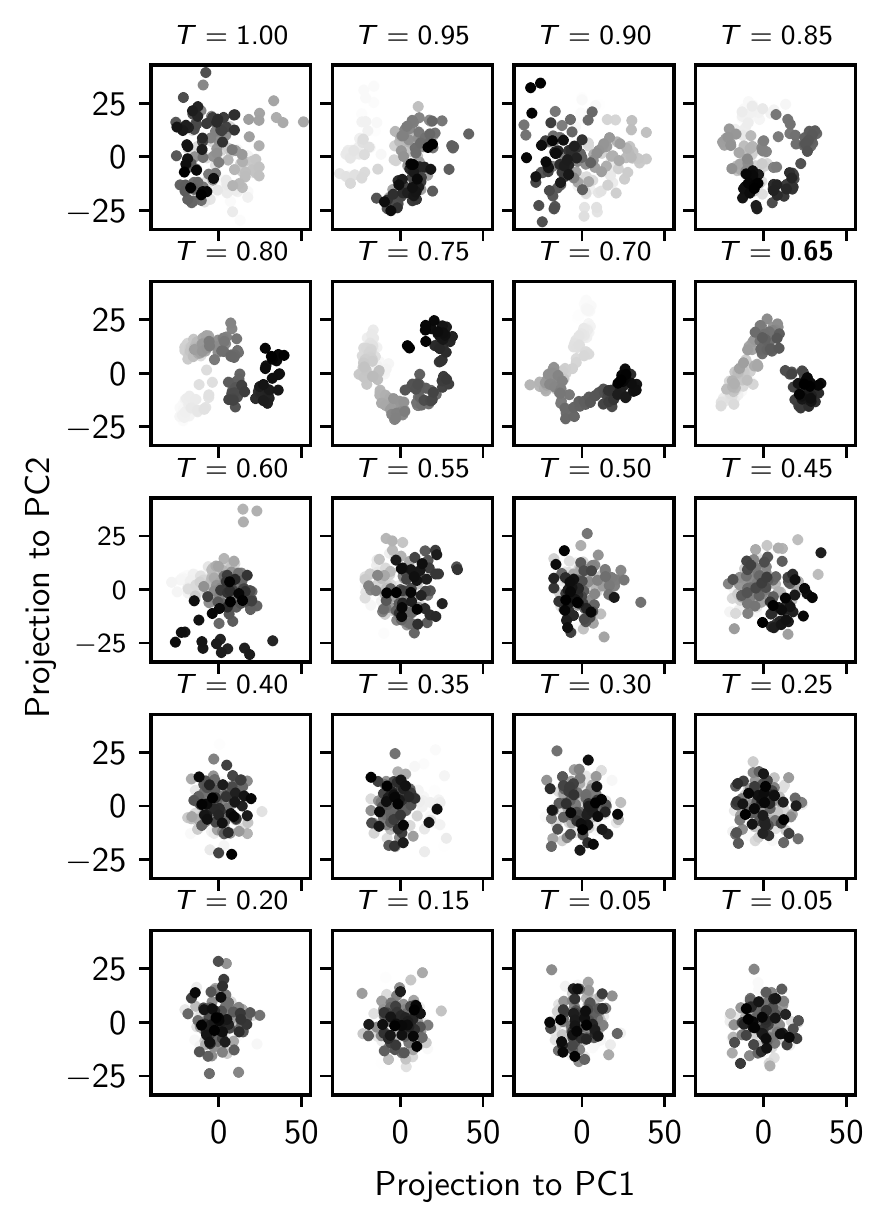}
	\singlespace\caption{ \changes{Temperature dependent PCA projections} for a randomly selected chain at each temperature $T$, as indicated.
	Color gradient for all projection correspond to the simulation time starting from light-gray to black. The fluctuations of data have the same magnitude after we standardise the input distances at each $T$. Hence, the PCA projections at $T \gg T_g$ and $T< T_g$ look visually similar. Around $T_g$, we see the change in the shape of the projections. This change is quantified using the first eigenvalue and the participation ratio as discussed in the main text. Data taken from the time window between $200\tau$ and $3\times 10^4\tau$ (gray area in \Cref{fig:MSD_timescale}). 
	}
	\label{fig:pca_temp}
\end{figure}

\clearpage



\label{si:sec:subchain}
\begin{figure}[!ht]
	\centering
\includegraphics[width=0.62\linewidth]
{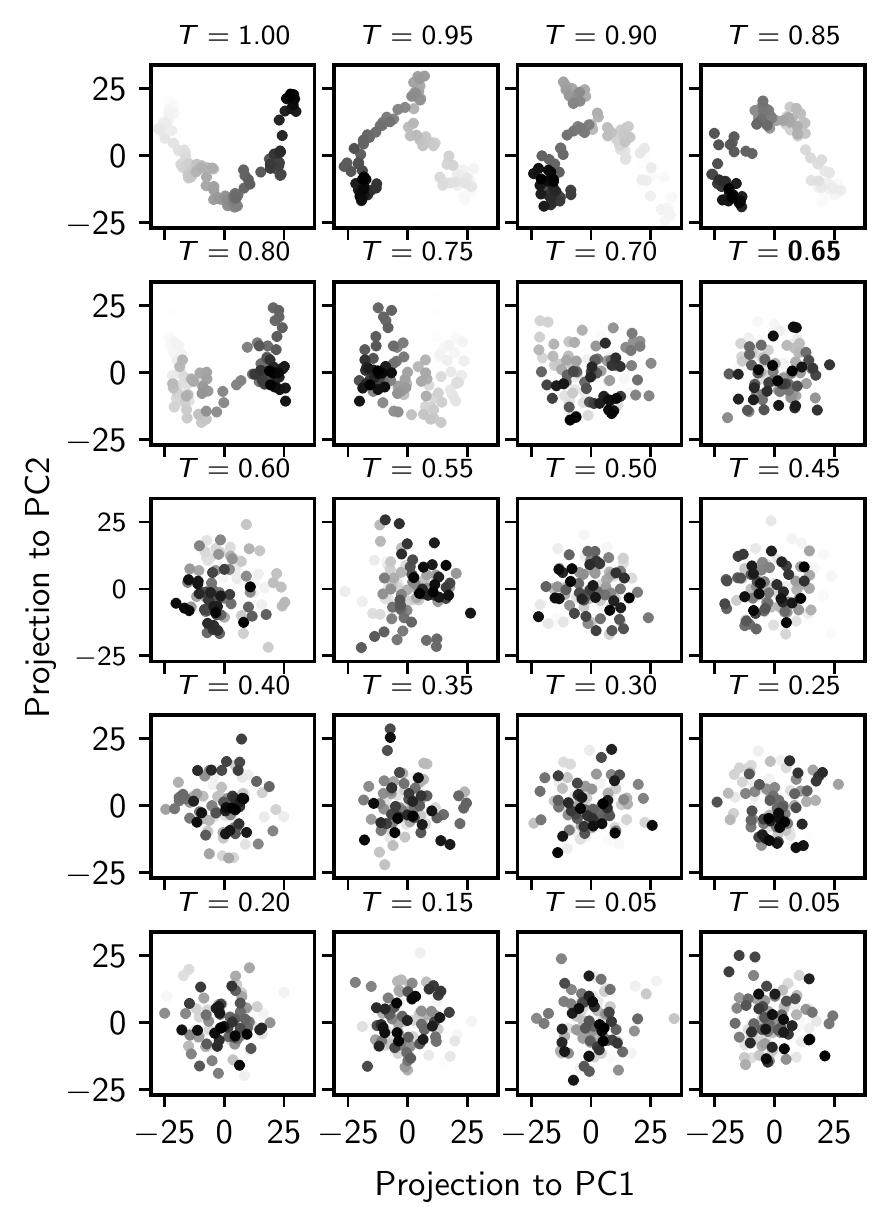}
\singlespace\caption{ \changes{Temperature dependent PCA projections} for a randomly selected chain at individual $T$ from a short MD time data (up to $20\tau$) (blue area in \Cref{fig:MSD_timescale}). No sharp jump in the first eigenvalue or PR around $T_g\approx 0.65\epsilon/k_B$ is observed.}
		\label{fig:short_time_all}
\end{figure}

%
\clearpage

\section{Results with reduced number of descriptors}
\label{si:sec:skip_monomers}
 \begin{figure}[h!]
	\centering
	\subfigure[]{
\includegraphics[width=0.36\linewidth]{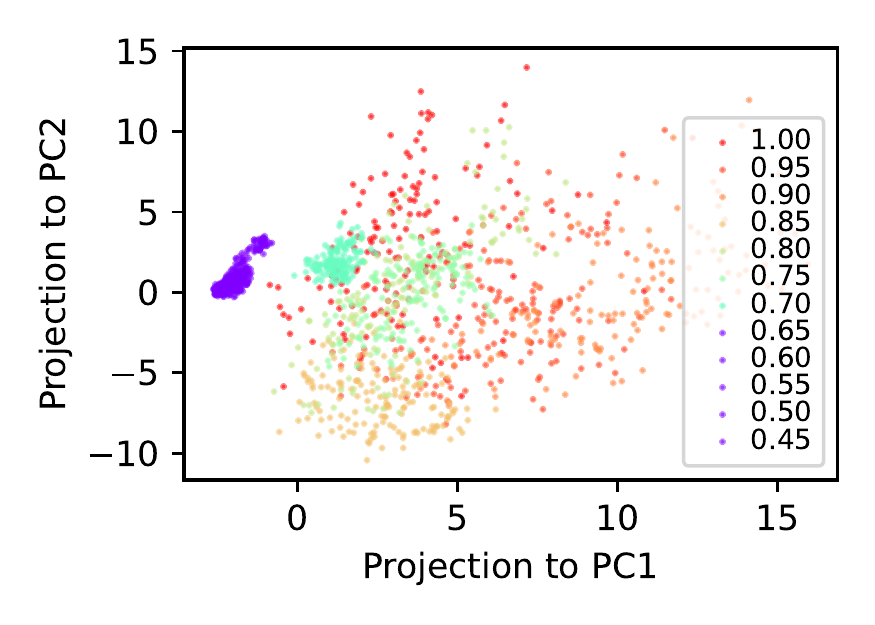}}
	\subfigure[]{
\includegraphics[width=0.32\linewidth]{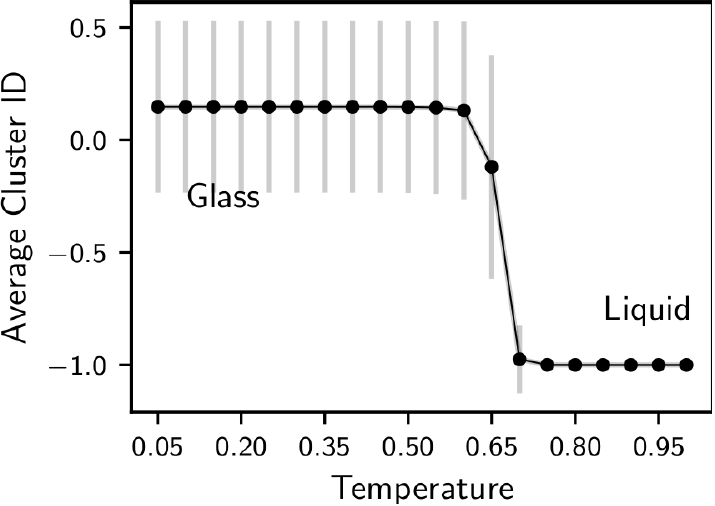}}
	\subfigure[]{
\includegraphics[width=0.32\linewidth]{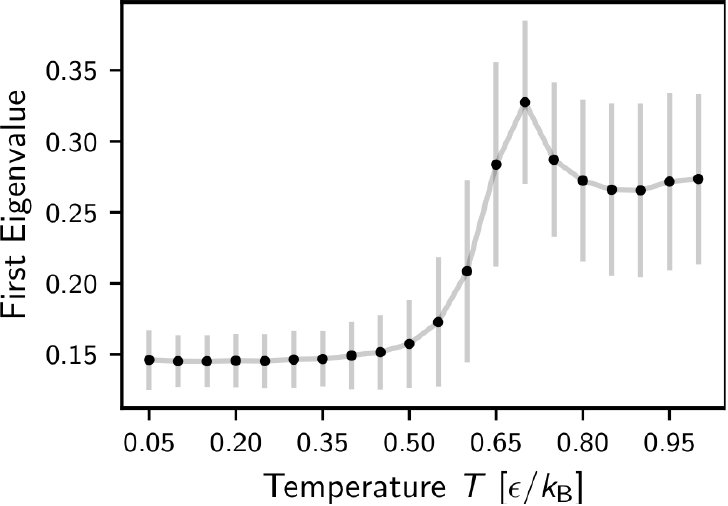}}
	\subfigure[]{
\includegraphics[width=0.32\linewidth]{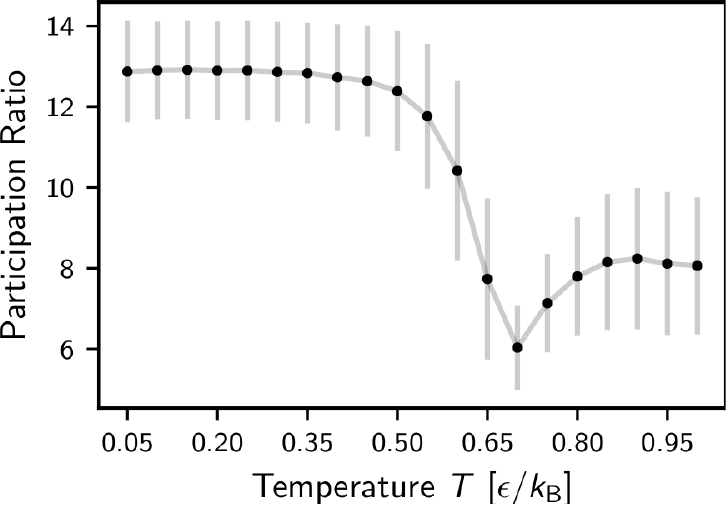}}
\singlespace\caption{The same analysis as in the main text done for reduced number of input descriptors (intra-chain distances calculated excluding four consecutive monomers).  (a) PCA projection of the same selected chain as given in Figure~1b of main text. (b) Average cluster IDs  $\langle n (T) \rangle$ as in Figure~1d. The first  eigenvalue (c) and PR (d) from individual temperature analysis (compare to Figure~3). 
}
\label{fig:SI-skipped-chain}
\end{figure}
 In the main text, we use the input data matrix $\mathbf{X}_{c} \in \mathbb{R}^{M \times L}$, where $L$ is the number of descriptors (the intra-chain  monomer-monomer distances of a single chain), and $M$ is the number of observations ($M=3000$ for Method I and $M=150$ for Method II). To reduce the short-range and highly correlated features we skip $\Delta_{m}$ consecutive monomers such that only $\lfloor \frac{n_m}{\Delta_{m}+1} \rfloor$ monomers with monomer indices $i\in \{k(\Delta_{m}+1)+1:k
 \;\textrm{is an integer with}\; k=0,1,\ldots,\lfloor \frac{n_m}{\Delta_{m}+1} \rfloor -1 \}$ from $n_m$ monomers in each chain are selected.
E.g. for $n_m=50$ and $\Delta_{m}=4$ the monomers with indices $i \in \{1,6,...,41,46\}$ are selected. 
 Our new descriptor space ($L = 10\times(10-1)/2$ = 45, $M$ remains the same) is relatively lowedimensional compared to the original ($L=1225$). In \Cref{fig:SI-skipped-chain}, we plot the single chain PCA projection, 
cluster IDs averaged over all chains, the first eigenvalue and PR from individual temperature analysis (as described in the main text) versus the temperatures $T$. All results are similar in nature after reducing the input feature space by excluding the contributions from the monomer pairs having chemical distances less than $(\Delta_{m}+1)$ 
along identical chains (data taken from the gray area of \Cref{fig:MSD_timescale}).  \changes{Nonetheless, to avoid the discussion on how many monomers one can skip for each specific system and make the description of our method more general we present the data with all intra-chain distances in the main text as PCA accounts for the highly correlated distances.}

\begin{figure}[!ht]
	\centering
 	\subfigure[]{
\includegraphics[width=0.33\linewidth]{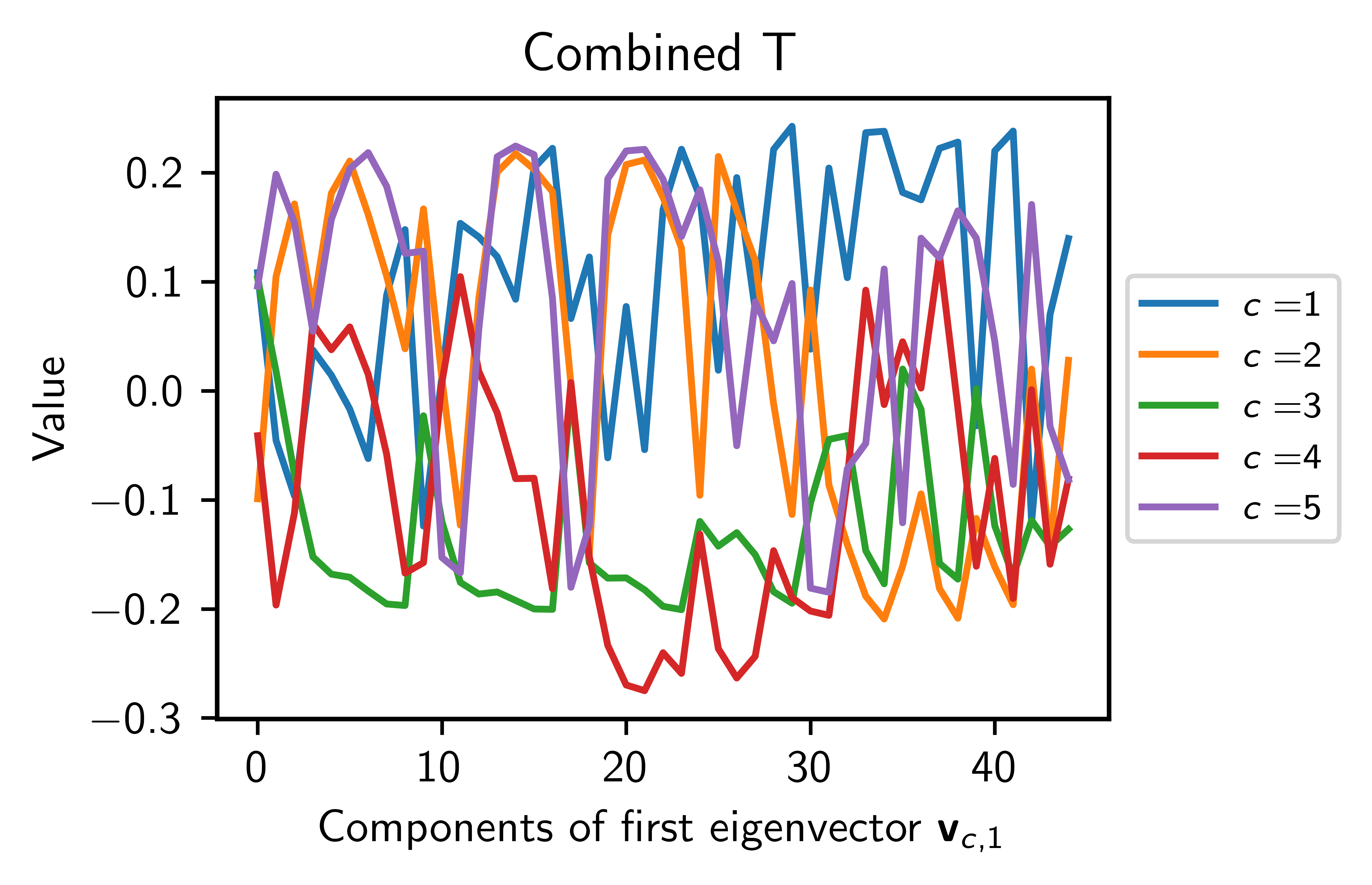}}
 	\subfigure[]{
\includegraphics[width=0.31\linewidth]{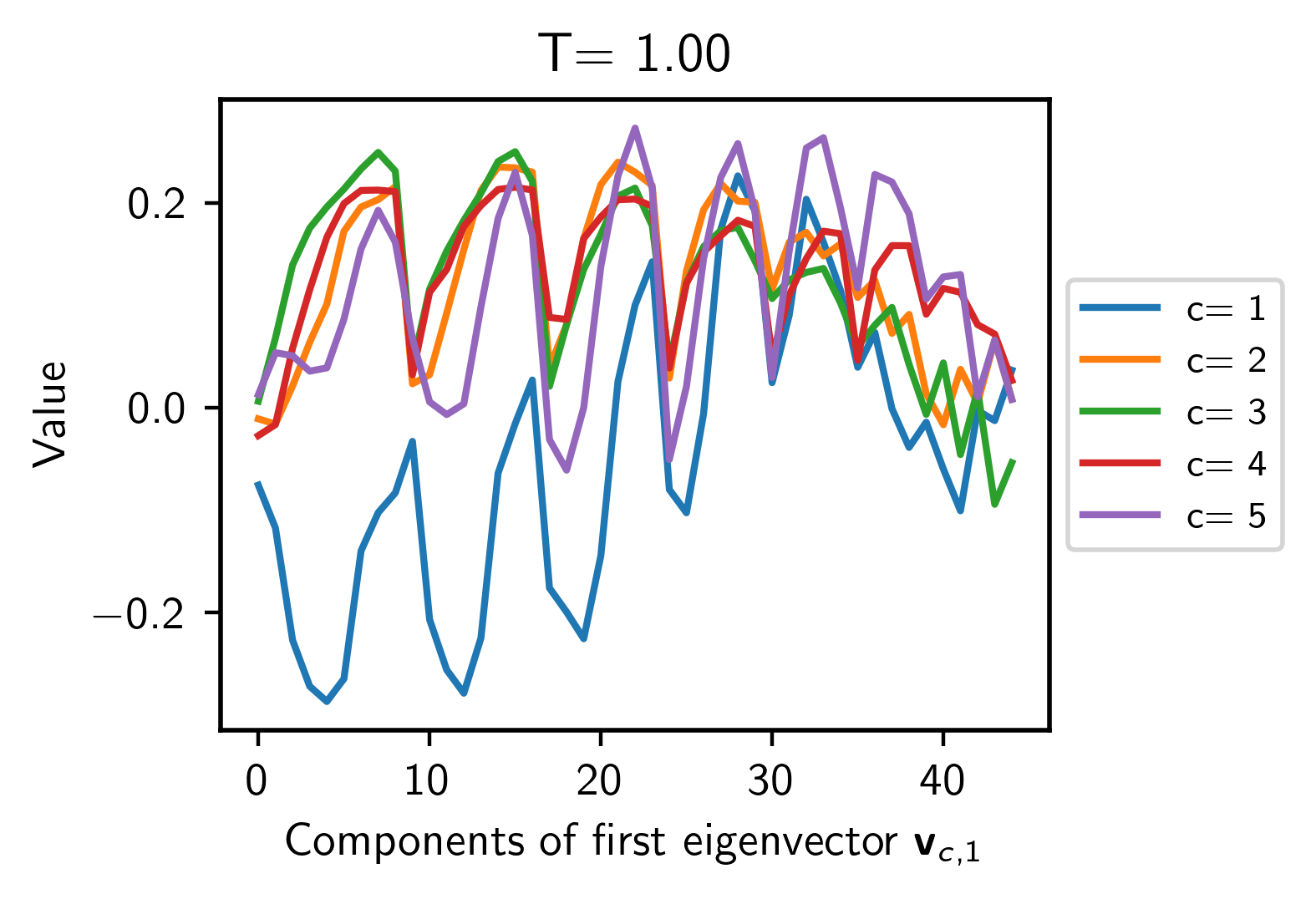}}
 	\subfigure[]{
\includegraphics[width=0.31\linewidth]{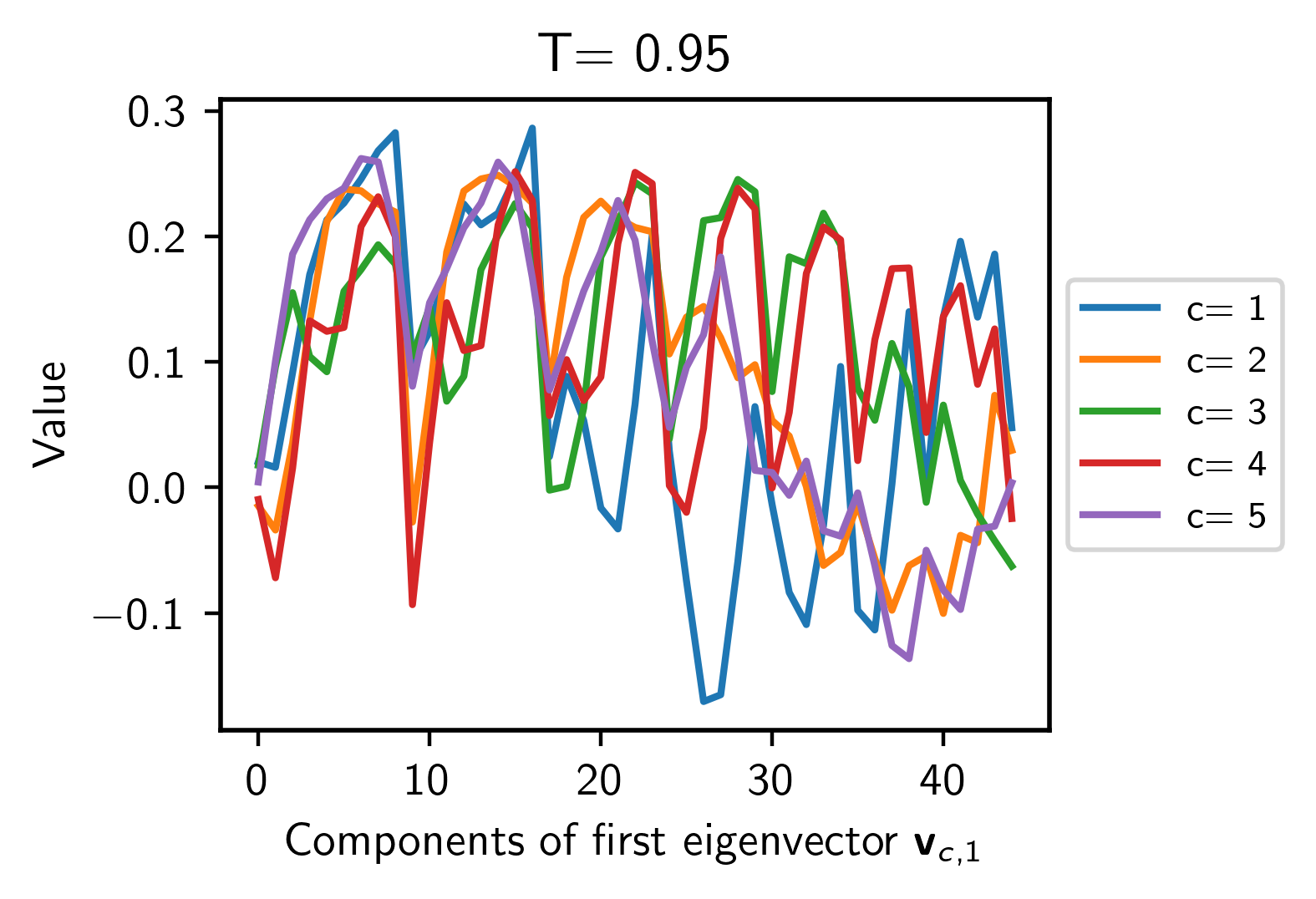}}
 	\subfigure[]{
\includegraphics[width=0.31\linewidth]{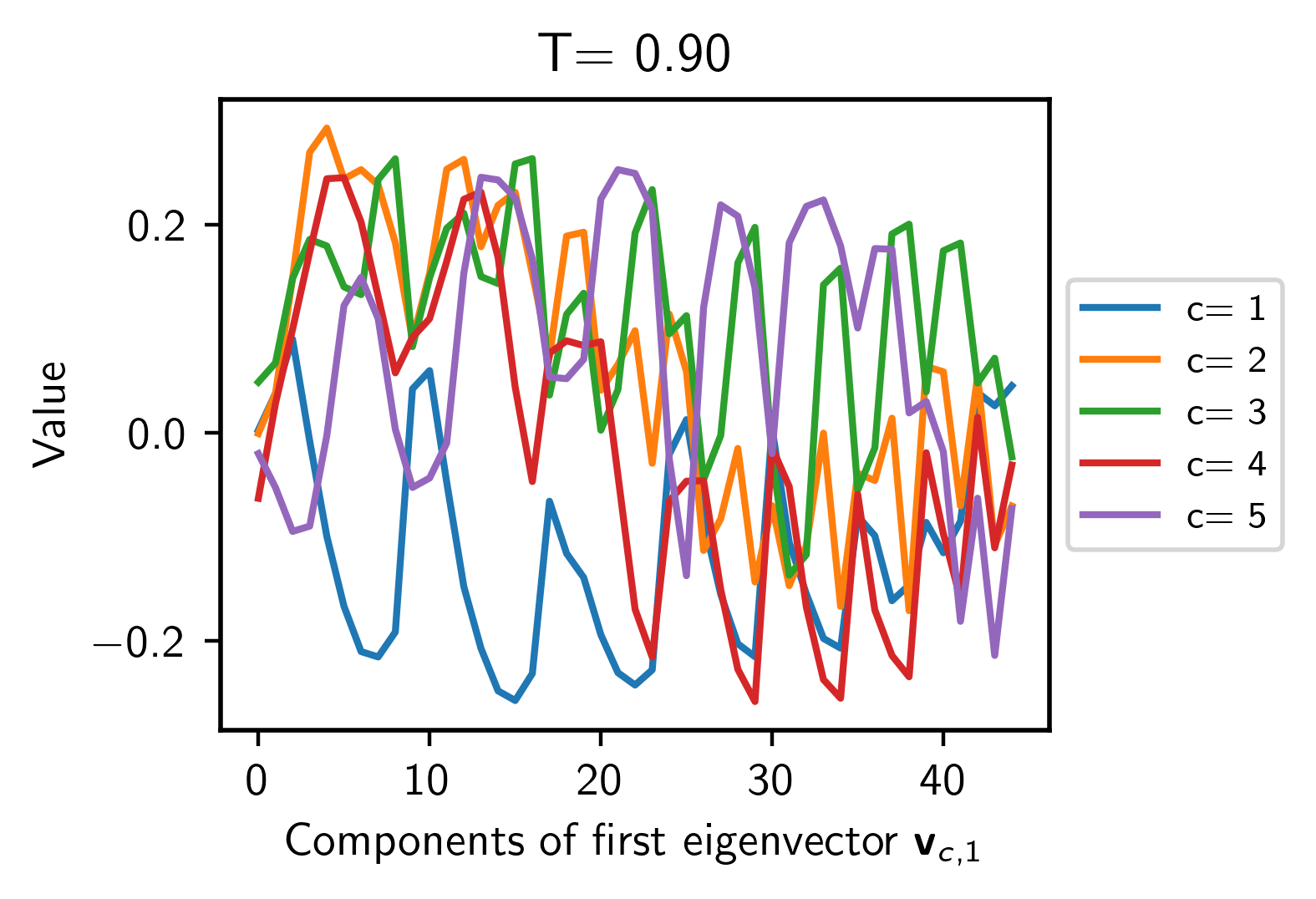}}
 	\subfigure[]{
\includegraphics[width=0.31\linewidth]{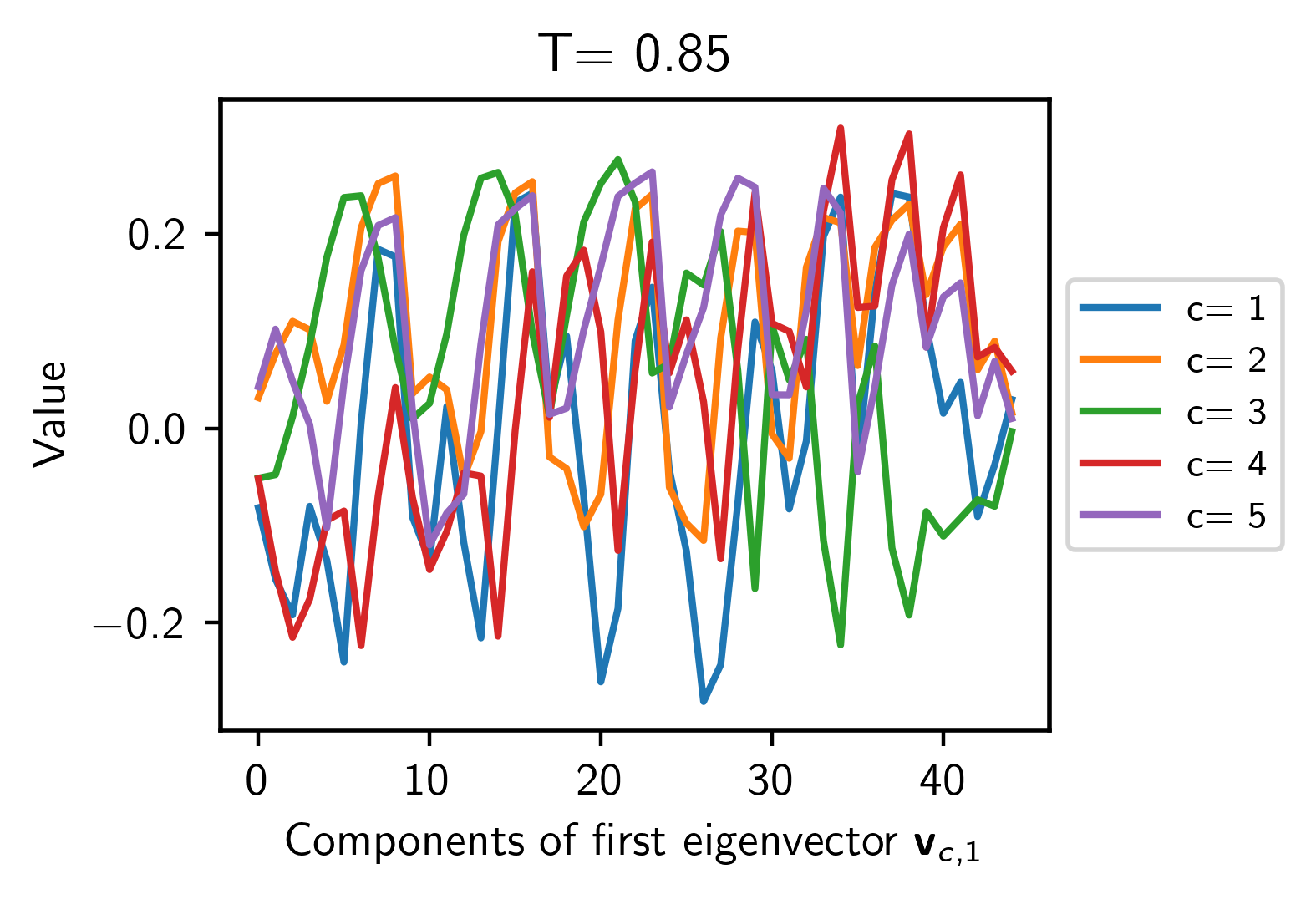}}
 	\subfigure[]{
\includegraphics[width=0.31\linewidth]{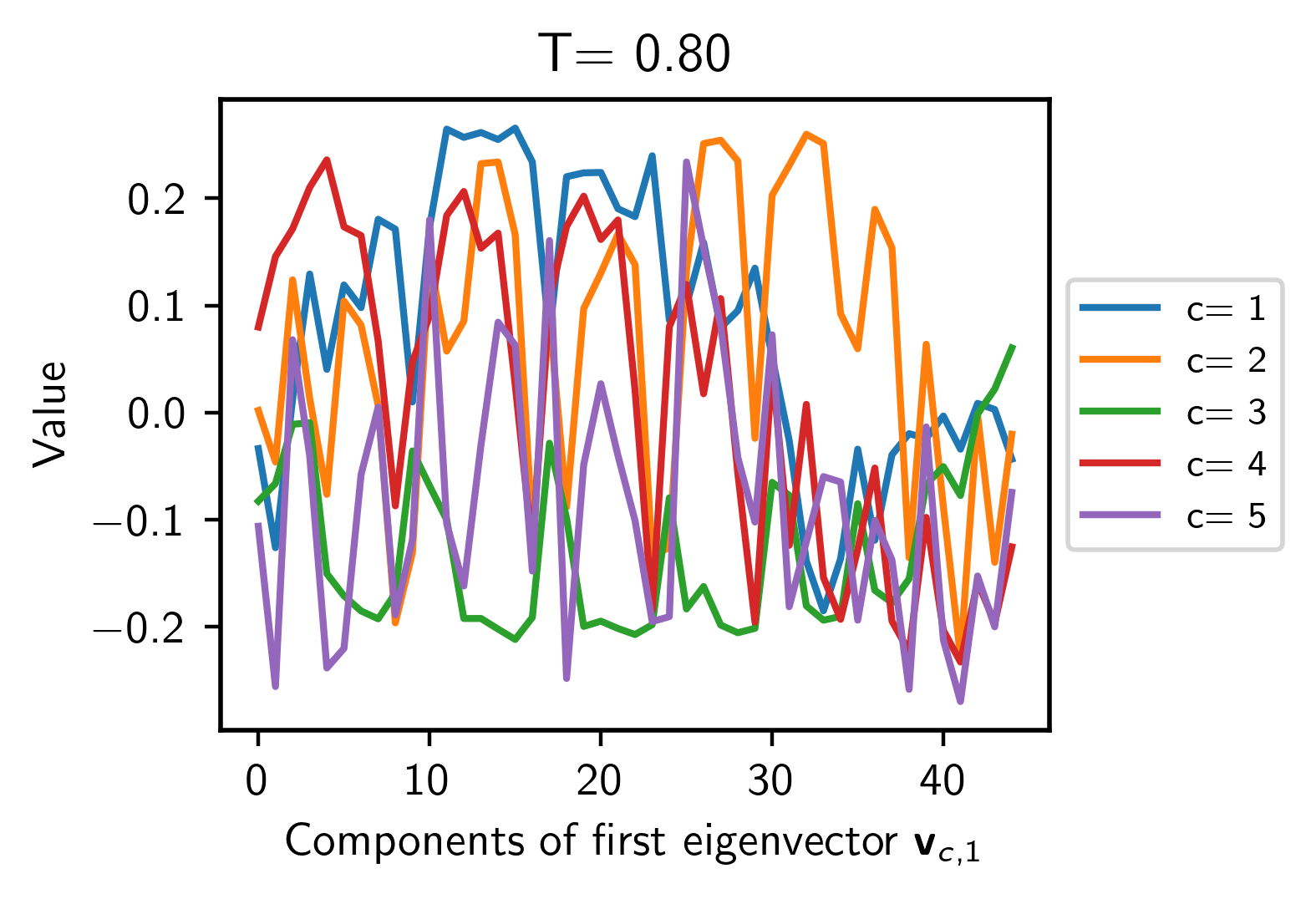}}
 	\subfigure[]{
\includegraphics[width=0.31\linewidth]{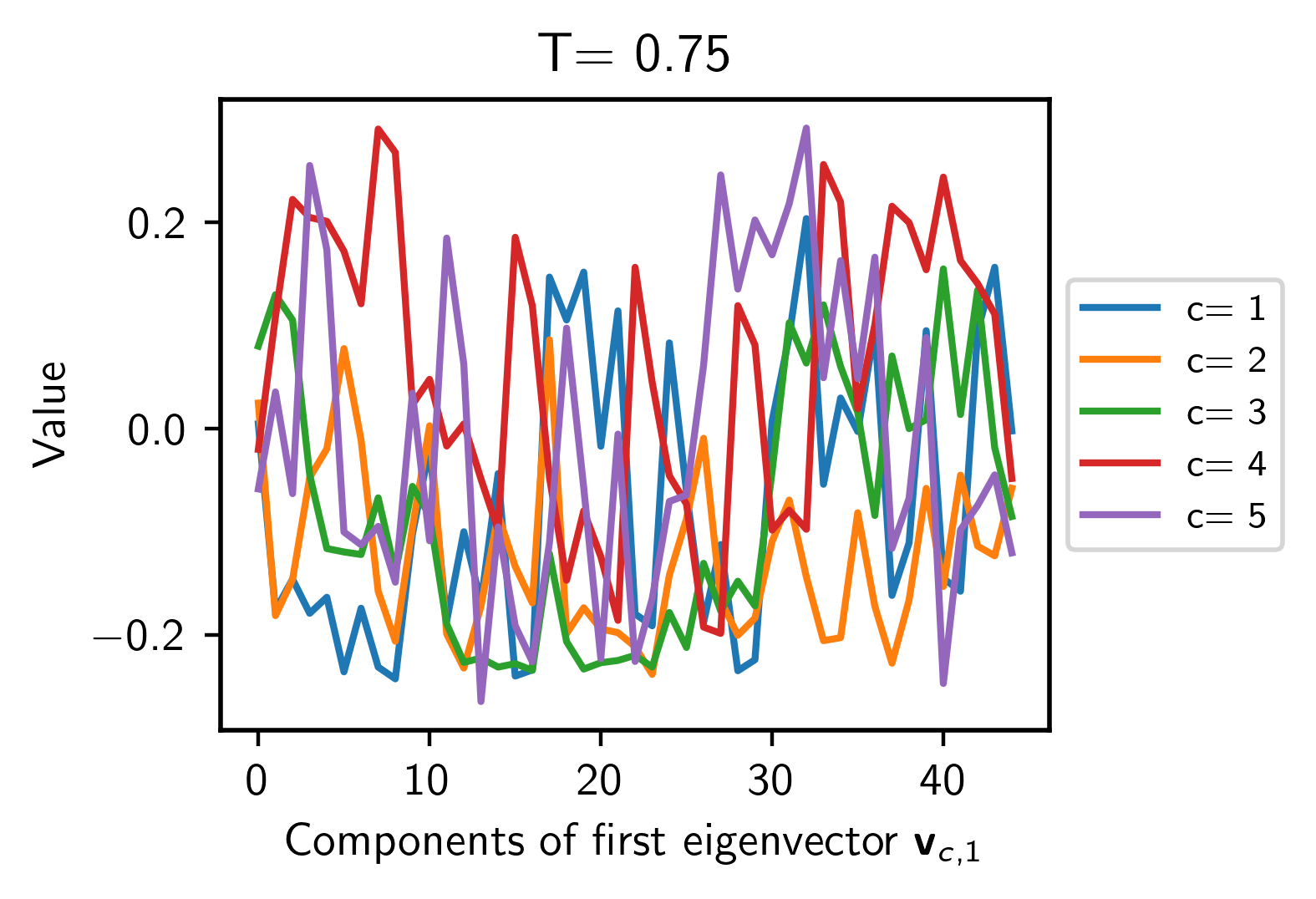}}
 	\subfigure[]{
\includegraphics[width=0.31\linewidth]{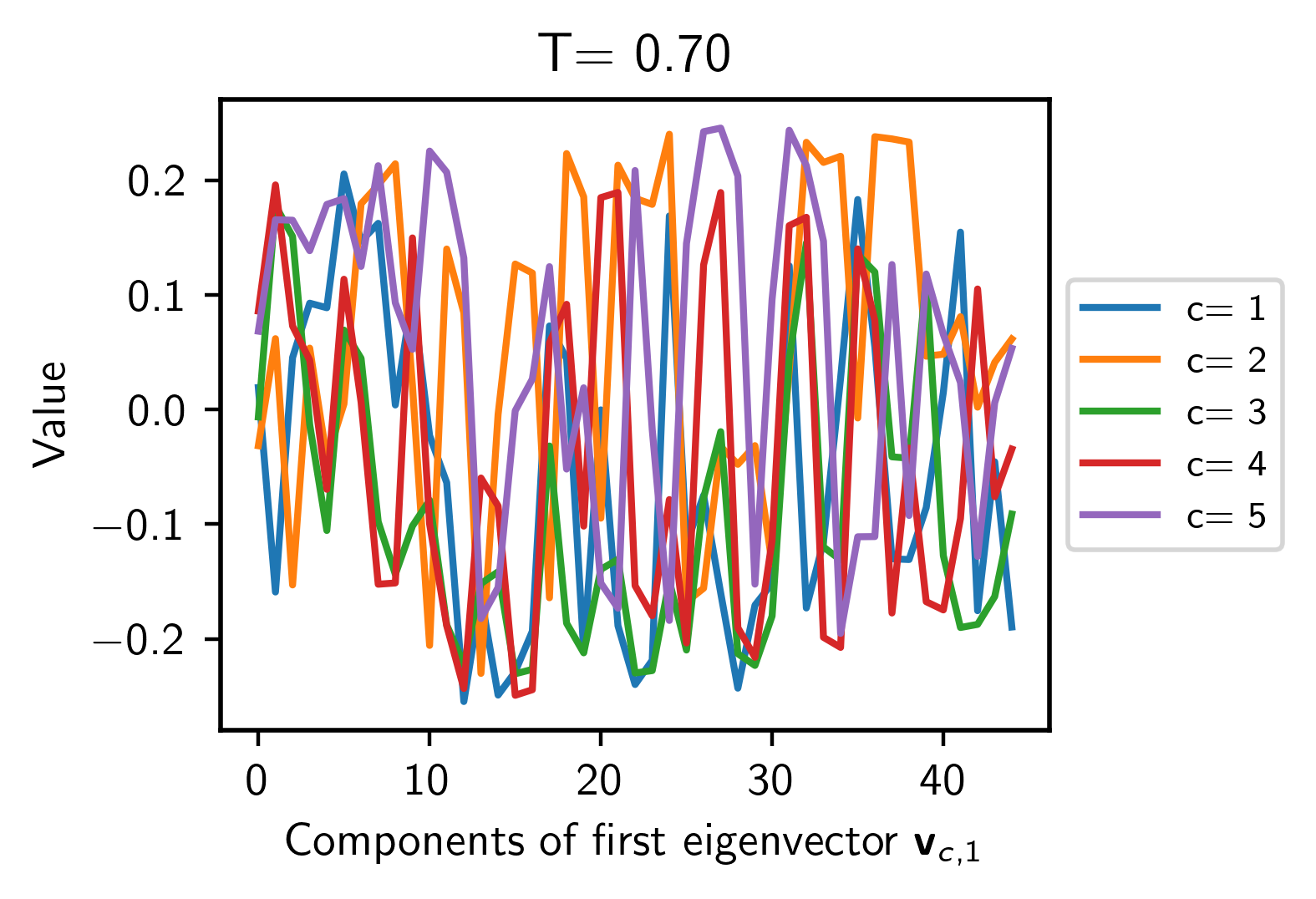}}
 	\subfigure[]{
\includegraphics[width=0.31\linewidth]{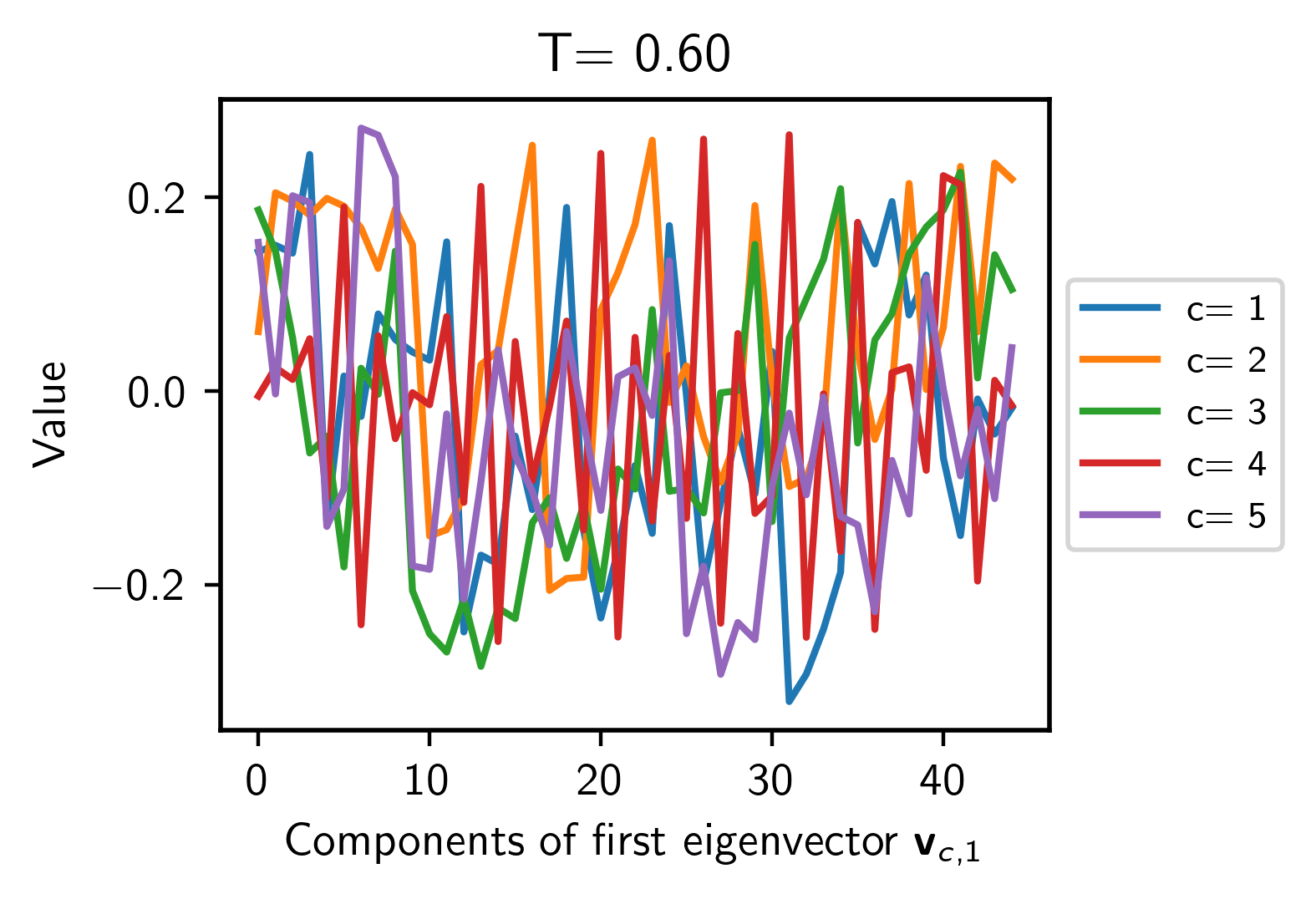}}

\caption{First eigenvectors ($\mathbf{v}_{c,1} \in \mathbb{R}^{45}$) of PCA applied independently to four randomly selected chains (on subset of 45 internal distances): (a) for Method~I (combined temperatures); (b)-(c) Method~II: different selected temperatures $T [\epsilon/k_B]= 1.00, \cdots, 0.60$ respectively. The x-axis shows the index of components of the first eigenvector ($\mathbf{v}_{c,1}=[v_1 \quad v_2 \quad \cdots \quad v_{45}]$) and the y-axis shows the value of each component. The chains in all plots are the same and indexed as $c =1,2,3,4,5$.}
\label{fig:eigenvectors}
\end{figure}

It is important to note the distinction between the eigenvectors' spaces used for Method~I and II. In the first case we have one set of $\mathbf{v}_{c,k}$, $k=1,...,P$ over all temperatures. For Method~II there are different independent sets of eigenvectors for each $T$ $\mathbf{v}_{c,k}(T)$. The examples of the first eigenvectors for four different chains are shown in \Cref{fig:eigenvectors} for (a) Method~I (combined temperatures) (b-i) Method~II (individual temperatures). For combined temperatures as well as for lower individual temperatures there are no common pattern/peaks for different chains. However, for the high temperatures (\Cref{fig:eigenvectors}b-e) we observe the similarity in the first eigenvectors, which suggests the importance of chemical distances at around 30 for the PCA projections obtained at this temperatures. The same importance was found for the PCA projections of the Method~I (see \Cref{fig:PC_hig_corr}c) and requires further study.

\end{document}